\documentclass[pra,twocolumn,showpacs,superscriptaddress,floatfi, nofootinbib]{revtex4-2}

\usepackage[english]{babel}
\usepackage[normalem]{ulem}

\usepackage{graphicx}
\usepackage{amsmath,amssymb,mathrsfs,esint}
\usepackage{multirow}
\usepackage{algorithm}
\usepackage[noend]{algpseudocode}
\usepackage{comment}
\usepackage{physics}
\usepackage{bbold}
\usepackage{tabularx}
\usepackage{subcaption}

\usepackage{bm}
\usepackage{relsize}

\usepackage[normalem]{ulem}
\usepackage[bookmarks=true,
   colorlinks=true,
   linkcolor=blue,
   urlcolor=blue,
   citecolor=blue,
   bookmarks=true,
   hyperindex=true
]{hyperref}

\usepackage{tikz}
\usetikzlibrary{quantikz2}

\DeclareUnicodeCharacter{00A0}{~}

\begin{document}

\title{Analysis of the action of conventional trapped-ion entangling gates in qudit space}

\author{Pavel A. Kamenskikh}\email{kamenskikh.pa@gmail.com}
\affiliation{P.N. Lebedev Physical Institute of the Russian Academy of Sciences, Moscow 119991, Russia}

\author{Nikita V. Semenin}
\affiliation{P.N. Lebedev Physical Institute of the Russian Academy of Sciences, Moscow 119991, Russia}

\author{Ilia V. Zalivako}
\affiliation{P.N. Lebedev Physical Institute of the Russian Academy of Sciences, Moscow 119991, Russia}

\author{\textcolor{black}{Vasiliy N. Smirnov} }
\affiliation{P.N. Lebedev Physical Institute of the Russian Academy of Sciences, Moscow 119991, Russia}

\author{Ilya A. Semerikov}
\affiliation{P.N. Lebedev Physical Institute of the Russian Academy of Sciences, Moscow 119991, Russia}
\affiliation{Russian Quantum Center, Skolkovo, Moscow 121205, Russia}

\author{Ksenia Yu.~Khabarova}
\affiliation{P.N. Lebedev Physical Institute of the Russian Academy of Sciences, Moscow 119991, Russia}
\affiliation{Russian Quantum Center, Skolkovo, Moscow 121205, Russia}

\author{Nikolay N.~Kolachevsky}
\affiliation{P.N. Lebedev Physical Institute of the Russian Academy of Sciences, Moscow 119991, Russia}
\affiliation{Russian Quantum Center, Skolkovo, Moscow 121205, Russia}

\begin{abstract}
Qudits, or multi-level quantum information carriers, present a promising path for scaling quantum computers. However, their use introduces increased complexity in quantum logic, necessitating careful control of relative phases between different qudit levels. In trapped-ion systems, entangling operations accumulate phases on specific levels that are no longer global, unlike in qubit architectures. Furthermore, the structure of multi-level gates becomes increasingly intricate with higher-dimensional Hilbert spaces.
This work explores the theory of these additional local and nonlocal phases, accumulated in M\o lmer--S\o rensen and Light-shift gates. We propose methods to actively compensate for these phases, enhance gate robustness against parameter fluctuations, and simplify native gates for more efficient circuit decomposition. Our results pave the way toward the practical and scalable implementation of qudit-based quantum processors.
\end{abstract}

\maketitle 

\section{Introduction}
Quantum computers have the potential to solve a variety of computational problems more efficiently than classical computers. This includes problems such as large-number factorization~\cite{shor1994Algorithms}, searching through large datasets~\cite{grover1996fast}, solving systems of linear equations~\cite{harrow2009Quantum} and others. There are different physical platforms on which a quantum computer can be built, and the most prominent ones are superconducting circuits~\cite{wu2021Strong, arute2019Quantum}, photons~\cite{zhong2020Quantum, madsen2022Quantum}, neutral atoms~\cite{ebadi2021Quantum, henriet2020Quantum, bluvstein2024Logical}, trapped ions~\cite{debnath2016Demonstration, zhang2017Observation, moses2023RaceTrack} \textcolor{black}{and semiconductor quantum dots~\cite{noiri2022Fast, xue2022Quantum}}. Trapped ions are among the first platforms proposed~\cite{monroe1995Demonstration, cirac1995Quantum} and have demonstrated some of the highest fidelities for single-qubit gates~\cite{smith2025SingleQubit} and two-qubit entangling gates~\cite{baldwin2021Highfidelity, loschnauer2024Scalable, hughes2025Trappedion}. However, the biggest challenge for real-world quantum devices based on trapped ions is scaling up the quantum register while maintaining the accuracy of quantum gates~\cite{bruzewicz2019Trappedion}. While it is possible to trap ion crystals with hundreds of particles~\cite{guo2024siteresolved}, the complexity of their control, especially of their motion, prevents their usage for universal digital computers. At the moment trapped-ion universal computers usually employ ion chains no longer than 30-40 particles~\cite{cheng2024Crosstalk, pogorelov2021Compact}. Several methods have been proposed to solve this problem, such as photon coupling between two ion traps~\cite{monroe2013Scaling, main2025Distributed} and trapped-ion QCCD (quantum charge-coupled device) architectures~\cite{kielpinski2002Architecture, pino2021Demonstration}.  

However, while increasing the number of qubits is the conventional approach to scaling quantum computers, it is not the only way to increase the dimension of the system's Hilbert space. An alternative is to harness a multi-level nature of ions and use more than two states in each particle for information encoding, transforming them from qubits to qudits~\cite{wang2020Qudits}. As each $d$-level qudit can be treated as $\log_2(d)$ qubits~\cite{kessel1999Multiqubit, kessel2002Implementation, nikolaeva2024Efficient, nikolaeva2024Universal} it is possible to increase an effective number of qubits in the system using the same number of ions with a rather small experimental overhead. Another feature of qudits is that additional states in trapped ions can be used as ancillary levels (ancillas) in quantum circuits, allowing for reduction of the number of gates required to perform an algorithm. For example, the Toffoli gate, used in the Grover algorithm~\cite{grover1996fast}, can be implemented using qudits, as described in Refs.~\cite{ralph2007Efficient, fedorov2012Implementation, nikolaeva2022Decomposing, nikolaeva2025Scalable} with a significantly smaller number of entangling gates. Quantum computation using qudits has already been demonstrated on several platforms~\cite{hill2021Realization, ringbauer2022universal, aksenov2023Realizing, kazmina2024Demonstration, zalivako2025multiqudit}, including systems based on trapped ions, such as~$^{40}\mathrm{Ca}^+$ ions~\cite{ringbauer2022universal},~$^{137}\mathrm{Ba}^+$ ions~\cite{low2025Control} and~$^{171}\mathrm{Yb}^+$ ions~\cite{aksenov2023Realizing, zalivako2025multiqudit, bian2026architecture}.

To fully exploit the potential of qudits in gate-based applications, it is important to have a universal set of native gates, enabling one to effectively decompose quantum circuits. The supported gate set significantly affects the number of operations required to perform a given circuit, influencing the fidelity of the algorithm. Although not that pronounced in the qubit case, it becomes more and more important as the qudit dimension increases. Qudit gates can influence differently various levels of the qudits and may require many parameters for their description depending on the experimental setup, complicating the transpilation process. One example is the entangling M\o lmer--S\o rensen (MS) gate~\cite{molmer1999Multiparticle}, which in the qubit case generates a~$\sigma_{X} \otimes \hat{\sigma}_{X}$ interaction between ions. The evolution operator corresponding to this gate also contains terms which are usually ignored as they result only in additional global phase acquired by qubits. However, when this gate is applied to qudits, they can no longer be ignored~\cite{ringbauer2022universal, nikolaeva2025Scalable}, as well as the Stark shift imposed on spectator states (qudit states not directly involved in the gate). 

In this paper, we provide a comprehensive theoretical analysis of the most common types of entangling gates, namely, M\o lmer--S\o rensen (MS) and Light-shift (LS), acting on trapped-ion qudits. We study the effect of these gates in the extended qudit Hilbert space, which often results in additional local and/or nonlocal phases acquired by various ions' states. In contrast to previous works~\cite{ringbauer2022universal,hrmo2023Native} which take into account interaction with only one center-of-mass (COM) motional mode, we consider interaction with multiple motional modes in the general case of a modulated laser pulse~\cite{choi2014Optimal, blumel2021Poweroptimal}. It is important for control of long ion chains in universal quantum processors where closely-spaced radial modes are usually utilized for entanglement. We also study ways to simplify the structure of the qudit entangling gates to optimize the process of quantum circuits decomposition. In particular, we propose pulse-shaping techniques canceling local phases acquired by spectator states during MS gate or at least making them robust to the drift of experimental parameters. For the LS gate, we propose spin-echo methods to reduce the number of nonlocal phases accumulated on qudit levels. We show that the proposed technique can simplify the LS gate interaction to an effective 
two-qubit entanglement in the case of qubit embedding.
This reduction can significantly simplify the decomposition of a qubit algorithm into native qudit gates when embedded qubits are used. We note that a similar method was previously proposed in~\cite{hrmo2023Native}, however its applicability was limited to a case, where only one qudit state was experiencing a strong light shift or qudit dimension is limited to $d=3$. Methods presented here provide a general solution for this problem. These findings are important for development of a robust and effective universal qudit quantum processor, simplification of the circuits decomposition process as well as for optimization of the calibration procedures.

Our work is organized as follows. 
In Section~\ref{sec:qudit_phases}, we obtain the expressions for the qudit phases accumulated in both MS
and LS gates. In Section~\ref{sec:pulse_shaping}, we review pulse shaping methods for the MS gate that aim to make the qudit phases robust to the fluctuations of experimental parameters, such as fluctuations of laser power and motional mode frequency drifts. Section~\ref{sec:spin-echo_qudit} presents several spin-echo sequences for the LS
gate designed to reduce the number of distinct nonlocal phases acquired by different qudit basis states. We discuss and summarise our results in Section~\ref{sec:discussion}.
\section{General description of entangling gates acting on trapped-ion qudits}
\label{sec:qudit_phases}
\subsection{M\o lmer--S\o rensen gate}\label{sec:MS~gate}
\begin{figure}[t]
\includegraphics[width= 7 cm]{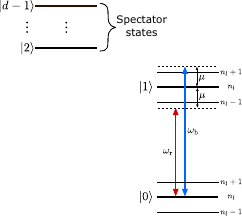}
\caption{Schematic representation of the M\o lmer--S\o rensen gate, acting on levels~$\ket{0}$ and~$\ket{1}$. Red and blue lines represent the spectral components of the bichromatic laser beam with frequencies $\omega_r$ and $\omega_b$. \textcolor{black}{The figure also shows the motional sidebands and indicates the phonon number~$n_l$ in the~$l^\mathrm{th}$ motional mode.}}
\label{fig:MSgate_scheme}
\end{figure}

First, we analyze the~$\hat{\sigma}_{X} \otimes \hat{\sigma}_{X}$~gate first proposed by M\o lmer and S\o rensen~\cite{molmer1999Multiparticle}. 
This gate involves a bichromatic electromagnetic field applied to a specific pair of ions. The bichromatic components interact with the red and blue secular motional sidebands of a transition between two qudit levels, as in Fig.~\ref{fig:MSgate_scheme}. Without loss of generality we consider our gate acting on levels~$\ket{0}$ and~$\ket{1}$. The field spectral components are symmetrically detuned by~$\mu$ from the carrier transition frequency, with~$\mu$ selected near the ions' motional mode frequencies. Making the Lamb-Dicke approximation~\cite{leibfried2003Quantum}, we obtain the Hamiltonian describing this interaction:
\begin{equation}
 \label{eq:MSgate}
 \hat{H}_{\mathrm{MS}} = \sum_{l, j} \hbar  \eta_{l, j}   g_{j}(t)  \left( \hat{a}_{l} e^{- i \omega_{l } t} + \hat{a}_{l}^{\dagger} e^{ i \omega_{l} t} \right) \hat{\sigma}^{(j)}_{\phi_{S}},
\end{equation} 
where \textcolor{black}{$g_j(t)$ is the pulse shape function for the $j^{\mathrm{th}}$ ion}, $\omega_l$ is the frequency of the $l^{\mathrm{th}}$ motional mode, $\eta_{l, j} $ is the Lamb-Dicke parameter characterizing the coupling of the $j^{\mathrm{th}}$ ion to the $l^{\mathrm{th}}$ mode, $\hat{a}_l$ and $\hat{a}_l^\dag$ are the annihilation and creation operators for the $l^{\mathrm{th}}$ mode. 
Phase~$\phi^{(j)}_{S}$ and operator~$\hat{\sigma}^{(j)}_{\phi_{S}}$ are defined as
\begin{equation}\label{eq:sigmaphi}
\hat{\sigma}^{(j)}_{\phi_{S}} = \hat{\sigma}^{(j)}_{X} \cos \phi_{S}^{(j)}  + \hat{\sigma}^{(j)}_{Y} \sin \phi_{S}^{(j)}, \quad \phi_{S}^{(j)} = \frac{\phi_\mathrm{r}^{(j)} + \phi_\mathrm{b}^{(j)}}{2},
\end{equation}
where~$\phi_{\mathrm{r,b}}^{(j)}$ is the initial phase of the red/blue components of the field for the~$j^{\mathrm{th}}$ ion, $\hat{\sigma}^{(j)}_X$ and $\hat{\sigma}^{(j)}_Y$ are the standard Pauli matrices that act in the two-level subspace of the~$j^{\mathrm{th}}$ qudit spanned by~$\ket{0}$ and~$\ket{1}$ (for example~$\hat{\sigma}_X^{(j)} = \ketbra{0_j}{1_j} + \ketbra{1_j}{0_j}$). The pulse function~$g_j(t)$ can in the general case be expressed in terms of the Rabi frequency~$\Omega_{j}(t)$ and the detuning~$\mu(t)$ (we further consider the detuning to be the same for all ions involved):
\begin{equation} \label{eq:g(t)}
g_j(t) = \Omega_j(t) \cos(\psi(t) + \phi_M^{(j)}), \quad \phi_M^{(j)} = \frac{\phi_\mathrm{r}^{(j)} - \phi_\mathrm{b}^{(j)}}{2},
\end{equation}
with the time-dependent phase $\psi(t) = \int_0^t \mu\left(t^{\prime}\right) d t^{\prime}$.

The laser fields induce additional Stark shifts on all levels: 
\begin{equation}
\label{eq:ham_MS_AC}
    \hat{H}_{\mathrm{AC}} =  \sum_{s = 0}^{d-1} \sum_{j} \hbar \delta_{j,s}^{\mathrm{AC}}(t) \ketbra{s_j}{s_j}.
\end{equation}
The full Hamiltonian~$\hat{H}$ consists of the MS gate term from Eq.~(1) and the Stark shift~(4), so~$\hat{H} = \hat{H}_{\mathrm{AC}} + \hat{H}_\mathrm{MS}$.
In practice, the Stark shifts for states~$\ket{0}$ and~$\ket{1}$ are usually calibrated to be equal:~$\delta_{j,0}^{\mathrm{AC}}(t)\equiv\delta_{j,1}^{\mathrm{AC}}(t)$, otherwise the nonzero differential AC stark shift between these states will cause significant errors~\cite{wu2018Noise}.
This compensation can be done, for example, by adding another off-resonant laser spectral component~\cite{franke2023quantum} or by introducing an imbalance in red and blue sideband amplitudes~\cite{kirchmair2009deterministic}. For the qudit case, these shifts, despite being equal, no longer represent a global phase and, therefore, need to be accounted for.
If~$\delta_{j,0}^{\mathrm{AC}}(t)\equiv\delta_{j,1}^{\mathrm{AC}}(t)$, the Hamiltonians~$\hat{H}_{\mathrm{AC}}$ and~$\hat{H}_\mathrm{MS}$ commute, and the evolution operator can be factorized. The complete proof is provided in Appendix~\ref{sec:appendix_Stark}. In the following, only the MS gate Hamiltonian~$\hat{H}_\mathrm{MS}$ is considered, and the effect of~$\hat{H}_{\mathrm{AC}}$ is discussed at the end of this subsection.

Phases and laser intensity for target ions can be calibrated so that~$\Omega_j = \Omega$ and~$\phi_{S}^{(j)} = \phi_{M}^{(j)} = 0$ for $j=1,2$~\cite{gerster2022Experimental}, which leads to \textcolor{black}{the same pulse function for both ions}~$g_j(t) = g(t)$ and~$ \hat{\sigma}^{(j)}_{\phi_{S}} = \hat{\sigma}^{(j)}_X$. This type of operation is called the~$\hat{\sigma}_{X} \otimes \hat{\sigma}_{X}$ gate or MS~gate. Its evolution operator can be obtained using the Magnus expansion \cite{blanes2009Magnus, magnus1954exponential}:
\begin{equation}\label{eq:evolution_MS}
\hat{U}_{\mathrm{MS}}' =  \exp(\hat{A}_1 + \hat{A}_{2}),
\end{equation}
where the first operator~$\hat{A}_1$ has the form
\begin{equation} \label{eq:A1}
\hat{A}_1 = \sum\limits_{l,j}(\alpha_{l,j}\hat{a}^{\dagger}_l-\alpha^*_{l,j}\hat{a}_l)\hat{\sigma}^{(j)}_{X},
\end{equation}
the complex number~$\alpha_{l, j}$ defines the position of the~$j^{\mathrm{th}}$ ion in phase space of the~$l^{\mathrm{th}}$ motional mode and can be written as
\begin{equation} \label{eq:alpha}
\alpha_{l,j}=-i \eta_{l,j}\int\limits_0^{\tau} g(t) e^{i \omega_l t}\,dt.
\end{equation}
In order to achieve ideal entanglement between ions, it is necessary to decouple the target ions from all modes, which means that
\begin{equation}
\alpha_{l, j} = 0\quad \forall l,j.
\label{eq:mode_decoupling}
\end{equation}
Under this condition~$\hat{A}_1 = 0$ at the end of the gate. This can be achieved by a proper choice of pulse function $g_j(t)$ and gate duration (this procedure is called pulse-shaping and is discussed in more detail in Appendix~\ref{sec:appendix_AWG_shaping}). The second operator~$\hat{A}_2$ in~\eqref{eq:evolution_MS} generates the entanglement between the states of the ions: 
\begin{equation}
    \label{eq:A2}
 \hat{A}_2 = i \sum\limits_{j,k = 1}^2\chi^{\mathrm{MS}}_{jk}\hat{\sigma}^{(j)}_{X}\hat{\sigma}^{(k)}_{X}.
\end{equation}
The expression for~$\chi^{\mathrm{MS}}_{jk}$ is given by
\begin{multline}
    \label{eq:chi_jk_MS}
\chi_{jk}^{\mathrm{MS}}=\sum\limits_l  \eta_{l, j}  \eta_{l, k} \int\limits_0^{\tau}dt_1\int\limits_0^{t_1} dt_2 \\
g(t_1) g(t_2) \sin(\omega_l(t_1-t_2)).
\end{multline}
Additional details on the Magnus expansion are provided in Appendix~\ref{sec:appendix_Magnus}. The summation in Eq.~\eqref{eq:A2} can be expanded as
\begin{multline}
    \label{eq:A2_expanded}
\hat{A}_{2} =  i \left[\chi^{\mathrm{MS}}_{11} \left( \hat{\sigma}_{X}^{(1)} \right)^2 \otimes \hat{\mathbb{1}}_2 + \chi^{\mathrm{MS}}_{22}  \hat{\mathbb{1}}_1 \otimes \left( \hat{\sigma}_{X}^{(2)} \right)^2 \right] + \\ + i  (\chi^{\mathrm{MS}}_{12} + \chi^{\mathrm{MS}}_{21}) \hat{\sigma}_{X}^{(1)} \hat{\sigma}_{X}^{(2)} , 
\end{multline} 
where~$\left( \hat{\sigma}_{X}^{(j)} \right)^2 = \ketbra{0_{j}}{0_{j}} + \ketbra{1_{j}}{1_{j}}$ projects onto the addressed qudit state subspace of the~$j^{\mathrm{th}}$ ion. For qudits, unlike qubits,~$\left( \hat{\sigma}_{X}^{(j)} \right)^2 \neq \mathbb{1}$, therefore, the first term in Eq.~(11), which acts locally on each qudit, cannot be omitted. The evolution operator~\eqref{eq:evolution_MS} then takes the final form:
\begin{equation}
\label{eq:evolution_qudits}
\hat{U}_{\mathrm{MS}} =  \exp(2 i \chi^{\mathrm{MS}}_{12} \hat{\sigma}_{X}^{(1)} \hat{\sigma}_{X}^{(2)} + i  \sum_{j = 1}^2\hat{\Phi}_j).
\end{equation}
The first term generates the entanglement between the target ions, so here we call $\chi^{\mathrm{MS}}_{12}=\chi^{\mathrm{MS}}_{21}$ \textit{nonlocal phases}. To realize a maximally entangling gate~(in the qubit sense), the pulse function is usually chosen in such a way that the phase~$\chi^{\mathrm{MS}}_{12}=\chi^{\mathrm{MS}}_{21}=\pi/8$. In the second term the operator~$\hat{\Phi}_j$ changes the phases of the qudit states of the~$j^\mathrm{th}$ ion:
\begin{equation}
    \label{eq:Phi_j LS}
    \hat{\Phi}_j = \chi_{jj}^{\mathrm{MS}} \left( \ketbra{0_j}{0_j} + \ketbra{1_j}{1_j} \right) +  \sum_{s = 0}^{d-1} \phi^{\mathrm{AC}}_{j,s} \ketbra{s_j}{s_j},
\end{equation}
where~$\phi^{\mathrm{AC}}_{j,s}  = - \int_{0}^{\tau} dt \delta_{j,s}^{\mathrm{AC}}(t)$ has been added from the Stark shift term~$\hat{H}_{\mathrm{AC}}$. These phases we call \textit{local}, respectively.
The effect of accumulation of the local phases on the qubit states~$\ket{0}$ and~$\ket{1}$ was first observed in Ref.~\cite{ringbauer2022universal}. There, the authors analyzed the special case of the MS gate in which the bichromatic field couples to a single center-of-mass motional mode, resulting in identical phases for both ions. Here we note that, in general, these local phases differ across ions when motional modes other than the center-of-mass are excited, as follows from~\eqref{eq:chi_jk_MS}. This is due to the different Lamb-Dicke factors for various ions in other modes.

\subsection{Light-shift gate}\label{sec:LS_gate}
\begin{figure}[t]
\subfloat[\label{fig:Walking wave}]
{\includegraphics[width = 8 cm]{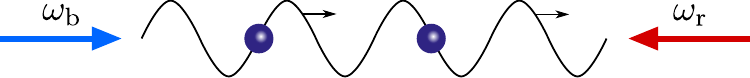}}
\hfil
\subfloat[\label{fig:LSgate_scheme}]{\includegraphics[width = 8 cm]{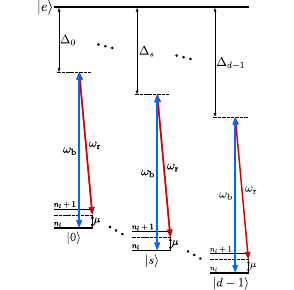}}
\caption{(a) An exemplary configuration of the LS gate, where two counter-propagating laser beams with frequencies~$\omega_\mathrm{r}$ and~$\omega_\mathrm{b}$ form a walking wave. (b) In this example state-depended Stark shifts on qudit levels arise from their off-resonant coupling to a single auxiliary level $\ket{e}$ by these laser fields. The frequency difference between two beams is chosen close to the secular frequency ($\omega_\mathrm{r}-\omega_\mathrm{b}=\mu\approx\omega_l$) to excite motion. Several motional levels with different phonon numbers $n_l$ are also depicted. \textcolor{black}{The qudit states are numbered here by their proximity to the auxiliary state.}}
\end{figure}

Another common type of entangling gate for ions is a Light-Shift~(LS) gate, also known as a geometric phase gate, providing a~$\hat{\sigma}_Z \otimes \hat{\sigma}_Z$ interaction between particles. This gate is implemented using a pair of overlapping beams with different frequencies that create a so-called walking wave, as shown in Fig.~\ref{fig:Walking wave}. The walking wave produces a light-induced Stark shift that depends on the state of the ions and on its position in the wave. As the interference pattern in the walking wave is moving, the ponderomotive force exerted on the ion from this light shift oscillates at each point. This, in turn, couples the motional modes with the electronic states. At the end of the gate the motion is again decoupled from ions' internal states while they acquire an nonlocal geometric phase that depends on their initial state. The~$\hat{\sigma}_Z \otimes \hat{\sigma}_Z$ gate was originally demonstrated by Leibfried et al. in Ref. \cite{leibfried2003Experimental}.

In addition to the original scheme, there are several other configurations of the~$\hat{\sigma}_Z \otimes \hat{\sigma}_Z$ gate, as described in Refs. \cite{aolita2007Highfidelity,baldwin2021Highfidelity, guo2024siteresolved}. The effective Hamiltonian is the same in all of these schemes, so we consider the most general model. 

In the most general configuration of the LS gate one or several pairs of laser beams with the frequency difference $\mu(t)$  create a walking wave at the target ions' positions. The operation of the gate can be illustrated by the simple case where each qudit state is off-resonantly coupled to a single auxiliary excited state~$\ket{e}$ (Fig. \ref{fig:LSgate_scheme}). 
After adiabatic elimination~\cite{baldwin2021Highfidelity} of all states, except ones constituting a qudit, we obtain a Hamiltonian:
\begin{equation}
    \label{eq:ham_LS_full}
    \hat{H}(t) = \hat{H}_{\mathrm{AC}}(t)  + \hat{H}_{\mathrm{LS}}(t),
\end{equation}
where the first term describes the AC-Stark shift~$\hat{H}_{\mathrm{AC}} =  \sum_{s = 0}^{d-1} \sum_{j} \hbar \delta_{j,s}^{\mathrm{AC}}(t) \ketbra{s_j}{s_j}$ which causes phase shifts on all qudit states. The second term~$\hat{H}_{\mathrm{LS}}(t)$ describes entanglement and after the Lamb-Dicke approximation has the form
\begin{equation}
    \label{eq:LSgate}
    \hat{H}_{\mathrm{LS}}(t) = \sum_{s = 0}^{d-1} \sum_{l, j}\hbar \eta_{l,j} g_{s, j} (t)  \left(  \hat{a}_l e^{- i \omega_l  t} + \hat{a}_l  ^{\dagger} e^{ i \omega_l  t}\right)  \ketbra{s_j}{s_j},
\end{equation}
which closely resembles the MS gate Hamiltonian~\eqref{eq:MSgate}, with two crucial differences. The first one is the electronic operators, which in this case are diagonal in the computational basis. The second one is the expression for the
The pulse function~$g_{s, j} (t)$:
\begin{equation} \label{eq:g(t)_LS}
g_{s, j}(t) = \Omega_{s, j}(t) \cos(\psi(t)),
\end{equation}
where~$\psi(t) = \int_0^t \mu\left(t^{\prime}\right) d t^{\prime}$, and~$\mu(t)$ is the time-dependent frequency difference between the laser beams. The~$\Omega_{s, j}(t)$ is the light shift amplitude for a given state $\ket{s}$. For unification with Eq.~\eqref{eq:g(t)} we further refer to it as to an effective Rabi frequency. Similarly to the MS~gate, the laser power on different ions can be set in such a way that the effective Rabi frequencies are uniform across target ions, i.e.~$\Omega_{s, j} = \Omega_s$ and pulse function does not depend on the ion: $g_{s, j}(t) = g_s(t)$.

In contrast to the MS gate, the hamiltonians~$\hat{H}_{\mathrm{AC}}$ and~$ \hat{H}_{\mathrm{LS}}$ in Eq.~(14) \textit{always} commute, since both terms are diagonal in the computational basis, allowing the evolution operator to be factorized. We therefore obtain the evolution operator for the~$\hat{\sigma}_{Z} \otimes \hat{\sigma}_{Z}$-gate using the Magnus expansion. As in case of the~$\hat{\sigma}_X \otimes \hat{\sigma}_X$ gate, the first operator~$\hat{A}_1$ in the Magnus expansion can be set to zero by an appropriate choice of the pulse function~$g_s(t)$ using the same techniques. The evolution operator for~$ \hat{H}_{\mathrm{LS}}$ can then be written as
 \begin{equation}
\label{eq:evolution_LS}
\hat{U}_{\mathrm{LS}}' =\exp\left( i \sum_{s,s' = 0}^{d-1} \sum_{j, k = 1}^2 \chi_{jk;ss'}^{\mathrm{LS}} \ketbra{s_{j}}{s_{j}}\times \ketbra{s'_{k}}{s'_{k}}
\right),
\end{equation}
with the phase~$\chi_{jk;ss'}^{\mathrm{LS}}$:
\begin{multline}
    \label{eq:chijk_LS}
    \chi_{jk;ss'}^{\mathrm{LS}} = \sum_{l} \eta_{l, j} \eta_{l, k}
        \int\limits_{0}^{\tau} d t_1\int\limits_{0}^{t_1} dt_2\\ g_s(t_1) g_{s'}(t_2)  \sin(\omega_l (t_1-t_2)) . 
\end{multline}
Although the~$\hat{\sigma}_{Z} \otimes \hat{\sigma}_{Z}$ and $\hat{\sigma}_{X} \otimes \hat{\sigma}_{X}$ gates share many features, there are several important differences. The MS phases~$\chi_{jk}^\mathrm{MS}$ from Eq.~\eqref{eq:chi_jk_MS} depend only on the ion indices, whereas the LS phases~$\chi_{jk;ss'}^{\mathrm{LS}}$ depend also on the qudit state. To simplify Eq. \eqref{eq:evolution_LS}, we rearrange the qudit states in descending order of their light shift amplitudes, such that~$|\Omega_0| \geq |\Omega_{1}| \geq ... \geq |\Omega_{d-1}|$ and normalize them relative to the $\Omega_0$: $\Omega_s = \theta_s \Omega_0$, ~$|\theta_{s}|  \leq 1$.  Then, the phase from Eq.~\eqref{eq:chijk_LS} becomes
\begin{equation}
\label{eq:chi_LS_ss}
\chi_{jk;ss'}^{\mathrm{LS}} = \theta_s \theta_{s'} \chi_{jk}^{\mathrm{LS}},
\end{equation}
where~$\chi_{jk}^{\mathrm{LS}} \equiv \chi_{jk;00}^{\mathrm{LS}}$. The evolution operator for the full Hamiltonian (taking into account Stark shifts) is:
\begin{equation}\label{eq:evolution_LS_full}
\hat{U}_{\mathrm{LS}} = \exp\left( 2 i \chi_{12}^{\mathrm{LS}} \sum_{s,s' = 0}^{d-1} \theta_s \theta_{s'} \ketbra{ss'}{ss'} + i \sum_{j = 1}^2\hat{\Phi}_j
\right),
\end{equation}
The first part generates entanglement between ions by applying different nonlocal phases to various qudit pair states. The second term~$\hat{\Phi}_j$ describes the acquisition of local phases by states of the~$j^{\mathrm{th}}$ ion:
\begin{equation}
    \label{eq:Phi_j LS}
    \hat{\Phi}_j = \sum_{s = 0}^{d-1} (\chi_{jj}^{\mathrm{LS}}\theta_s^2 + \phi^{\mathrm{AC}}_{j,s}) \ketbra{s_j}{s_j},
\end{equation}
where~$\phi^{\mathrm{AC}}_{j,s}  = - \int_{0}^{\tau} dt \delta_{j,s}^{\mathrm{AC}}(t)$ is the phase due to the AC-Stark shift of state~$\ket{s}$. 

It is important also to consider a special case of LS gate, which we call a \textit{zero-order} LS gate, where~$|\theta_{s\geq 1}|~\ll~1$, meaning that only one of the states strongly interacts with the laser fields. This happens, for example, if the laser frequencies are close to some strong transition from the state~$\ket{0}$ which is not the case for other states or if, due to selection rules or a specific polarization choice, the condition~$\theta_s = 0$ for~$s \geq 1$ is enforced. This configuration of the LS gate for qudits was considered in Ref.~\cite{hrmo2023Native}, where~$\ket{0}$ is chosen from the~$S_{1/2}$ term of $^{40}\textrm{Ca}^+$, while the remaining qudit states are selected from the~$D_{5/2}$ term. As the polarizability of the $S_{1/2}$ term at 401~nm, which wavelength was used for LS gate implementation, is much higher than for a $D_{5/2}$ term, the zero-order condition was satisfied.

Under the above approximation, the evolution operator for the LS gate takes a simpler form:
\begin{equation}
\label{eq:evolution_LS_0}
    \hat{U}_{\mathrm{LS}}^{(0)} = \exp\left( 2 i \chi_{12}^{\mathrm{LS}} \ketbra{00}{00} + i \sum_j(\chi_{jj}^{\mathrm{LS}} + \phi_{j,0}^{\mathrm{AC}}) \ketbra{0_j}{0_j}
\right).
\end{equation}

Comparing the entangling terms in evolution operators of the MS and LS gates (equations~\eqref{eq:evolution_qudits} and~\eqref{eq:evolution_LS_full}, respectively), MS gate acts on the~$\ket{0}$ and~$\ket{1}$ states only, thereby providing entanglement only in the subspace spanned by these states in both qudits. In contrast, the entangling term of the LS gate in general case acts simultaneously on all qudit states, making the structure of the entanglement much more complicated. For a more detailed analysis of the MS and LS gates in application to qudit-based quantum computing architectures, see~\ref{sec:discussion}.

Both the MS and the zero-order LS evolution operators~\eqref{eq:evolution_qudits} and~\eqref{eq:evolution_LS_0} respectively have only one nonlocal phase, so it may seem that these gates are locally equivalent, that is, they can be transformed into one another by only using single-qudit local gates. However, it can be shown~\cite{semenin2026equivalence} that they are, in fact, not locally equivalent for~$d>2$, unless~$d=3$ and~$2\chi_{12}^{\mathrm{MS}} = 2\chi_{12}^{\mathrm{LS}} = \pi$. This means that even the simplest case of the LS gate~\eqref{eq:evolution_LS_full} is generally not reducible to the MS gate if the gates are extended to qudits. Therefore, the qudit MS and LS gates are qualitatively different.

\section{Robustness to fluctuations of experimental parameters}
\label{sec:pulse_shaping}
Any trapped-ion entangling gate has a certain degree of sensitivity to fluctuations in the parameters of the experimental setup. Among the most prominent effects reducing the gate fidelity are laser power and phase fluctuations, nonzero temperature of the ion chain and the heating thereof, as well as secular frequency drifts.
The laser phase fluctuations and temperature have essentially the same effect for qudits as for qubits.
Namely, both MS and LS gates are insensitive to the initial temperature in the first order, while the MS gate is much more susceptible to the laser phase noise than the LS gate due to its resonant nature.

Another source of error which can arise due to fluctuations in parameters is the residual spin-motion entanglement, determined by the values of~$\alpha_{l,j}$ in~\eqref{eq:alpha}. For the MS gate, it is evident that this error is independent of the qudit dimension~$d$, since the spectator states are not addressed by the laser field, so they cause no change in the motional state. For the LS gate, strictly speaking, the number of different displacement parameters is equal to the qudit dimension, as is seen from the sum over~$s$ in~\eqref{eq:LSgate}, making the qudits generally more sensitive to this error than qubits. However, since the displacement parameters for different~$s$, being created by the same bichromatic field, are proportional to each other, they can all be set to zero simultaneously by pulse shaping of~$g_s(t)$. We mentioned this earlier, before obtaining the evolution operator~\eqref{eq:evolution_LS}. Furthermore, an optimized shaped pulse has the same form for qubits and qudits, only varying in magnitude for different~$s$. In a more practical case, where the number of nonzero~$g_s(t)$ does not increase with the qudit dimension (i.e. the zero-order LS), the error is independent of~$d$, making qudits and qubits equally sensitive to residual spin-motion entanglement.

It is important to note that the Hamiltonians~\eqref{eq:MSgate} and~\eqref{eq:LSgate} are obtained in the Lamb-Dicke approximation, which keeps only the terms linear in~$\eta_{l,j}$, while neglecting the terms of greater orders, as well as the carrier term. The latter originates either from the interaction of the bichromatic field with the carrier~$\ket{0}\leftrightarrow\ket{1}$ transition (in the MS gate) or from the time-dependent Stark shift of each qudit state induced by the beatnote between the bichromatic components (in the LS gate). These neglected terms can also cause additional errors.

The influence of the terms described above has been studied in~\cite{wang2022Fast} for the MS gate, but the analysis can be extended to the LS gate in a straightforward way.
The carrier term leads to a state-dependent phase shift of the states addressed by the gate, while the high-order Lamb-Dicke terms cause, in the leading order, a number of state-dependent cross-mode couplings, which induce additional spin-motion entanglement. The effect of these terms is the same for both qubits and qudits in the MS gate, since only the states~$\ket{0}$ and~$\ket{1}$ are addressed. The argument for the LS gate is identical to the one provided for spin-motion entanglement. Namely, a properly chosen pulse~\cite{wang2022Fast} can set the carrier and high-order errors to zero for all~$s$ simultaneously, and the shape of the pulse is the same for qubits and qudits. Therefore, assuming a shaped pulse that compensates for these errors in qubits, qudits are equally as sensitive to them in both MS and LS gates.

Changes in secular frequencies and in laser power result in a more complicated evolution in qudit systems. The reason for this is the high number of local and nonlocal phases for qudits, dependent on both of these parameters. A qudit MS gate has one nonlocal phase~$\chi_{12}^{\mathrm{MS}}$ and two local phases~$\chi_{11}^{\mathrm{MS}}$, $\chi_{22}^{\mathrm{MS}}$ which are not global, unlike their qubit counterparts. A general LS gate has $d(d-1)/2$ nonlocal and $d -1$ local phases (here we take into account that one of the local phases can be neglected as global). While some of local phases can be canceled similarly to the qubit case by applying additional laser fields or by tuning the balance between spectral components, the general picture for the qudit case is less robust to the fluctuations of the experimental parameters.

In both the MS and LS gates, the phases~$\chi_{jk}^{\mathrm{MS}}$, $\chi_{jk;ss'}^{\mathrm{LS}}$  and~$\chi_{jj}^{\mathrm{MS}}$, $\chi_{jj;ss}^{\mathrm{LS}}$ are dependent on secular frequencies and, as such, are susceptible to their drifts. Moreover, these drifts also result in residual entanglement between electronic and motional degrees of freedom of ions at the end of the gate. The latter effect is manifested in the violation of the decoupling condition~\eqref{eq:mode_decoupling}. Therefore, the effect of frequency drifts on the gate fidelity is two-fold. Thus, a substantial effort must be made to improve the gate stability against secular frequency drifts. At the same time, changes in laser power only affect the aforementioned phases, not the decoupling condition~\eqref{eq:mode_decoupling}, as long as it is satisfied at the end of the gate.

One of the ways to improve the robustness of qudit gates is to adopt pulse-shaping techniques widely used in qubit processors~\cite{choi2014Optimal, schafer2018Fast}. They are used to reduce gates' sensitivity to the secular frequency drifts, which we further refer to as to their stabilization. Although the general type of the LS gate has too many phases to stabilize, it can still be done for the MS gate. It has only one nonlocal and two local phases dependent on secular frequencies. In the following subsection we show that multi-tone pulse shaping~\cite{blumel2021Poweroptimal} can be used to stabilize both local and nonlocal phases with respect to secular frequency drifts. We also show that $\chi_{jj}^{\mathrm{MS}}$ can be stabilized and set to zero at the same time. This allows one to make local phases acquired by some of the qudit states during the gate immune to not only secular frequency drifts but also to laser power fluctuations. 

\subsection*{Stabilization of the local and nonlocal phases in the MS gate against experimental parameters drifts using pulse shaping}

\begin{figure*}[hbt]
\centering
\includegraphics[width=0.98\linewidth]{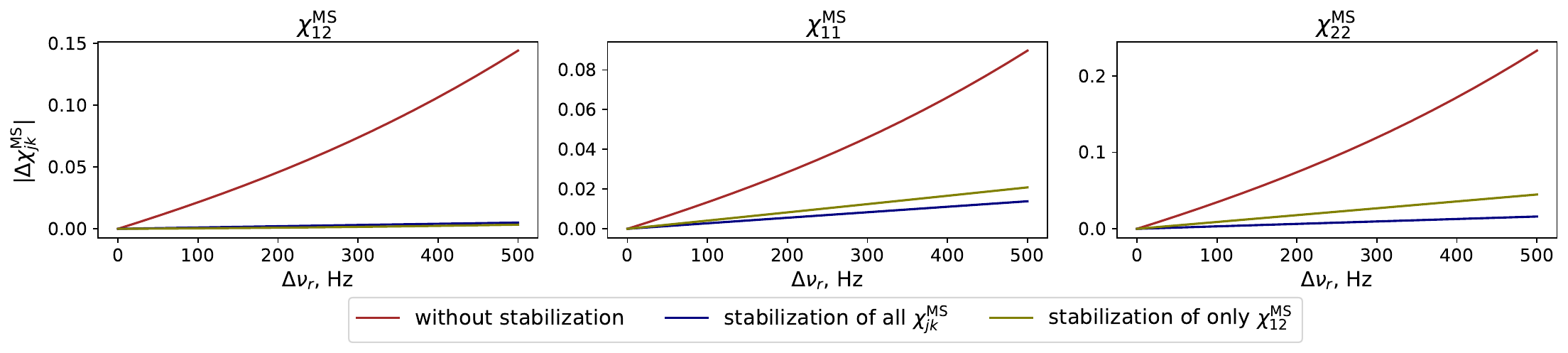}
\caption{Sensitivity of the phases~$\chi^{\mathrm{MS}}_{jk}$ to drifts of motional modes~$\omega_l$. Fluctuations are calculated for multi-tone MS~gate.}
\label{fig:delta_chis_fluctuations}
\end{figure*}

The multi-tone shaping technique discussed here simultaneously and continuously modulates the laser field's amplitude, frequency, and phase~\cite{haddadfarshi2016High, blumel2021Poweroptimal, shapira2020Theory, webb2018Resilient}. The basic idea behind this method is to expand the pulse function~$g(t)$ in a Fourier series:
\begin{equation}\label{eq:gFourier}
g(t)=\sum_{p=0}^{P-1} \Omega_p \sin( \mu_p t),
\end{equation} 
where~$\mu_p$ is the detuning of the~$p^{\mathrm{th}}$ tone. The real constant coefficients~$\Omega_p$ define a pulse comprising~$P$ tones with uniform frequency spacing. The straightforward way of experimental implementation of such a multi-tone MS gate is via an arbitrary waveform generator (AWG) \cite{blumel2021Poweroptimal}.

Introducing a vector~$\vec{\Omega}=(\Omega_0,...,\Omega_{P-1})$ we can express the  phases~$\chi_{jk}^{\mathrm{MS}}$ as:
\begin{equation}
    \label{eq:chi_AWG}
    \chi_{jk}^{\mathrm{MS}} = \Vec{\Omega}^T  \mathbf{S}_{jk}  \Vec{\Omega},
\end{equation}
where~$\mathbf{S}_{jk}$ is a symmetric matrix. To stabilize these phases against secular frequency drifts, we use the projection method~\cite{blumel2021Poweroptimal} which projects out $M$ eigenvectors of~$\partial\mathbf{S}_{jk}/\partial\omega_l$ with the highest eigenvalue magnitudes, which correspond to the lowest robustness. Furthermore, we use the moments method~\cite{blumel2021Poweroptimal} to stabilize the decoupling condition~\eqref{eq:mode_decoupling} against the aforementioned drifts as well. For a more detailed explanation of the pulse shaping and stabilization methods, see Appendix~\ref{sec:appendix_AWG_shaping}.

For an experimental implementation of the gate, shapes that require the least laser power are favorable, due to technical limitations of the setup. Moreover, lower power causes smaller errors from carrier coupling and crosstalk between neighboring ions. We therefore optimize pulse shapes by minimizing the second norm of the amplitude vector,~$\| \vec{\Omega} \|_2$, which is proportional to the mean laser power~\cite{james1998Quantum}:
\begin{equation}\label{eq:laserpower}
    \| \vec{\Omega} \|_2^2 = \sum_n \Omega_n^2 =   \vec{\Omega}^T \vec{\Omega}. 
\end{equation}
Overall, we state the optimization problem as follows:
\begin{equation}
    \label{eq:SLSQP_system}
    \begin{aligned}
    \min_{\vec{\Omega}}\quad &\| \vec{\Omega} \|^2_2\\
    \textrm{s.t.}\quad
        &\vec{\Omega}^T \mathbf{S}_{12}\vec{\Omega} = \chi,\\
     &\vec{\Omega}^T \mathbf{S}_{jj}\vec{\Omega} = 0, \quad j = 1, 2 \\
     &\mathbf{R}^T \vec{\Omega} = 0,
    \end{aligned}
\end{equation}
where the matrix~$R$ encompasses the stabilized decoupling condition and the projection method simultaneously.
We solve this optimization problem using the sequential least squares programming algorithm (SLSQP) from the SciPy library \cite{virtanen2020SciPy}. 

Using SLSQP we numerically optimize pulse shapes for the multi-tone MS gate without stabilization ($M = 0$) and with~$M = 5$, \textcolor{black}{where $M$ is the number of projected out eigenvectors of~$\partial\mathbf{S}_{jk}/\partial\omega_l$ for each mode.} 
It can be shown that a solution without stabilization exists for any parameter set using the multi-tone pulse-shaping method (see Appendix~\ref{sec:appendix_existence}). However, the usage of the projection method, used for stabilization, constrains the solution space. Consequently, for~$M \neq 0$, a stabilized pulse shape cannot always be guaranteed.
Nevertheless, we have verified that stabilized solutions do exist for a wide range of experimentally relevant conditions, including:
\begin{enumerate}
    \item Arbitrary ion pairs within a chains of 3 to 20 ions
    \item The secular frequencies typically employed in experiments:
        \begin{itemize}
            \item Radial,~$\nu_r \in [2, 5]$ MHz
            \item Axial,~$\nu_z \in [200, 700]$ kHz.
        \end{itemize}
\end{enumerate}
Figure \ref{fig:delta_chis_fluctuations} shows the simulated deviations
\textcolor{black}{
\begin{equation*}
|\Delta\chi_{jk}| = |\chi_{jk} - \chi_{jk}^{\mathrm{(target)}}|
\end{equation*}}
of the MS phases from their target values caused by the fluctuation in the radial center-of-mass secular frequency~$\nu_r$. 

The power-optimal shapes for this plot were obtained for \textcolor{black}{
the following set of parameters:
\begin{itemize}
    \item Secular frequencies~$\nu_r = 3.7$ MHz and~$\nu_z = 400 $ kHz;
    \item Ion numbers: $4^\mathrm{th}$ and~$6^\mathrm{th}$ in a 10-ion chain;
    \item Gate duration: $500 \ \mu \mathrm{s}$;
    \item Target entanglement (nonlocal) phase:~$\chi = \pi /8$;
    \item Number of Fourier harmonics: $P = 512$.
\end{itemize}
}
Stabilizing all three phases  $\chi_{12}^{\mathrm{MS}}$,  $\chi_{11}^{\mathrm{MS}}$ and  $\chi_{22}^{\mathrm{MS}}$ requires projecting out~$3\times M\times N$ vectors, \textcolor{black}{where~$N$ is the number of ions in the chain}. This decreases the number of degrees of freedom and, in turn, increases the power required during the gate.
For a more power-efficient approach, we make use of the fact that these phases are not completely independent, but connected through the Lamb-Dicke parameters (see Eq.~\eqref{eq:chi_jk_MS}). Therefore, stabilizing only one of them might already be sufficient. Green curves in
 Fig.~\ref{fig:delta_chis_fluctuations} show the same phase deviations as before, but in the case when only~$\chi_{12}^\mathrm{MS}$ is stabilized using the projection method. This reduces the number of projected vectors to~$N\times M$ and achieves performance comparable to full stabilization, making it more desirable for practical implementation.

\textcolor{black}{To evaluate the power needed to drive the pulses obtained, we calculate the largest Rabi frequency achieved during the gate~
$\Omega_{\max}$. Table~\ref{tab:powers} contains the comparison between different stabilization schemes in terms of this parameter.
\begin{table}[h]
    \centering
    \renewcommand{\arraystretch}{1.5} 
    \begin{tabular}{c|c}
    Stabilized phases~$\chi_{jk}^\mathrm{MS}$ & $\Omega_{\max}/(2\pi)$, kHz \\
        \hline
    None & 121\\
    $\chi_{12}^\mathrm{MS}$, $\chi_{11}^\mathrm{MS}$, $\chi_{22}^\mathrm{MS}$ & 431\\
    $\chi_{12}^\mathrm{MS}$ & 322
\end{tabular}
    \caption{\textcolor{black}{Largest Rabi frequencies during the gate versus stabilization methods}}
    \label{tab:powers}
\end{table}
}

Thus, conventional pulse-shaping methods developed for qubits also prove to be efficient for the MS gate acting in the qudit space in the wide range of experimental parameters.

\section{Reduction of the qudit LS gate with spin-echo sequences}\label{sec:spin-echo_qudit}
The evolution operator in Eq.~\eqref{eq:evolution_LS_full} contains two terms: the nonlocal term
\begin{equation}
    \label{eq:evolution_LS_entangle}
    \hat{U}_{\mathrm{LS}}^{\mathrm{ent}} = \exp\left( i \sum_{s,s' = 0}^{d-1} \phi_{ss'}^{\mathrm{LS}} \ketbra{ss'}{ss'}
\right),
\end{equation}
where~$\phi_{ss'}^{\mathrm{LS}} = 2 \chi_{12}^{\mathrm{LS}} \theta_s \theta_{s'}$, and the local term
\begin{equation}
    \label{eq:evolution_LS_single}
    \hat{U}_{\mathrm{LS}}^{\mathrm{single}} = \exp\left( i ( \hat{\Phi}_1 + \hat{\Phi}_2)
\right).
\end{equation}
\textcolor{black}{In this section we investigate only the nonlocal term. The local phases are considered separately in Section~\ref{sec:
non_entangling_echo}}

In general case it contains roughly~$d^2/2$ distinct nonlocal phases~$\phi_{ss'}^{\mathrm{LS}}$. This is quite undesirable in an experimental setup, because not only do their number grow quadratically with~$d$, they are also not completely independent and thus difficult to control and stabilize. Moreover, even if they were controllable, usage of such a "raw" LS gate would require significant computational resources for circuits transpiling due to large number of gate parameters and its complicated structure. Lastly, for an arbitrary set of LS phases, the application of the gate creates entanglement between all qudits' states. For some applications, it is beneficial to generate the maximal qudit entanglement in terms of, for example, the Schmidt rank~\cite{hrmo2023Native, brennen2005Efficient}. However, if the qudit states are used to \emph{embed} several qubits~\cite{fedorov2022Quantum, kiktenko2025Colloquium}, a high degree of entanglement between various states after gate application becomes a burden. The more
convenient
gate structure in this case would be the action on two particular
embedded qubits in different ions,
leaving the other, spectator, 
qubits unperturbed. Here we investigate approaches to simplify the structure of the qudit LS gate with spin-echo sequences, allowing one to achieve such a gate behavior. 

The spin-echo method is already widely used in LS gates~\cite{roos2008Ion, baldwin2021Highfidelity}. It works by splitting the whole gate
into several smaller gates in such a way that in total they create the target phase shift. These smaller gates are called loops because during each of them the motional states travel along a closed trajectory on the phase diagram, leaving motion and internal degrees of freedom decoupled at the end. Spin-echo pulses are inserted between the loops to cancel residual systematic errors and suppress slow experimental parameter noise~\cite{roos2008Ion}. 

\begin{figure}[h]
    \centering
    \includegraphics[scale=0.6]{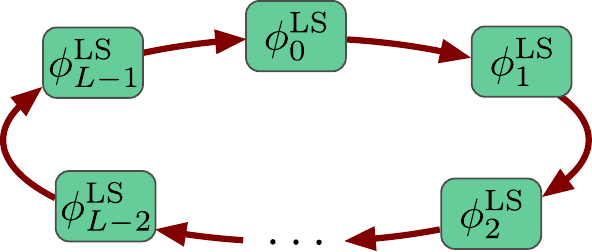}
    \caption{\textcolor{black}{A general spin-echo sequence of length~$L$. Each rectangle represents a qudit basis state, and the red arrows correspond to the echo pulses}}
    \label{fig:spin_echo_principle}
\end{figure}

The way spin-echo can help simplify the qudit LS gate structure can be understood as follows. During each LS gate loop, a specific basis state accumulates a certain phase. Each echo pulse transfers its population to another basis state. The subsequent LS gate loop adds another phase shift, according to this new state, and so on (see fig.~\ref{fig:spin_echo_principle}). 
In the end, the population is transferred back to the starting state, and the total phase it has acquired is equal to the sum of all nonlocal phases of the basis states that have been "visited" by it.

The goal we pursue here is to maximally simplify the structure of the gate, so the spin-echo sequence that we seek must satisfy the following condition: the total phase~$\phi_{ss'}^{\mathrm{LS}}$ in Eq.~\eqref{eq:evolution_LS_entangle} accumulated during the sequence by an initial state~$\ket{ss'}$ takes the least possible number of different values among all~$s,s'$. In this case, we refer to the resulting LS gate as the \emph{reduced} gate. In the following, we consider the reduction of the LS gate to two phases, which is the simplest non-trivial case. 

For the reduced gate with 2 phases, the entire two-qudit Hilbert space~$\mathbf{H}$ can be separated into two subspaces:
\begin{equation}
\label{eq:LS_H0_H1}
    \mathbf{H} = \mathbf{H}_0 \oplus \mathbf{H}_1.
\end{equation}
These spaces are spans of the basis vectors corresponding to the respective total acquired phases:
\begin{equation}
\label{eq:LS_symmetry}
    \phi_{ss'}^{\mathrm{LS}} = \begin{cases}
        \varphi_0, \quad &\mathrm{if} \ \ket{ss'} \in \mathbf{H}_{0}, \\
        \varphi_1, \quad &\mathrm{if}  \ \ket{ss'} \in \mathbf{H}_{1}.
    \end{cases}
\end{equation}
Obviously, for an arbitrary vector inside either of these spaces, the gate action would be the same as for the basis vectors that span them. Note that with a suitable choice of~$\mathbf{H}_0$ and~$\mathbf{H}_1$ the sequence, despite only featuring two phases, can still provide maximal qudit entanglement (in terms of Schmidt rank)~\cite{hrmo2023Native}.

In the following, we use the standard notation for native single-qudit rotations for trapped ions~\cite{ringbauer2022universal, aksenov2023Realizing}. The first type of gate performs a rotation of the state in the subspace spanned by the basis states~$\ket{j}$ and~$\ket{k}$ by angle $\theta$ around axis $\phi$ in the equatorial plane of the Bloch sphere:
\begin{equation}
    \hat{R}_{\phi}^{jk}(\theta) = \exp(- i \frac{\theta}{2}\hat{\sigma}_{\phi}^{jk}).
\end{equation}
Here~$\hat{\sigma}_{\phi}^{jk}$ is the Pauli matrix~\eqref{eq:sigmaphi} in the subspace spanned by~$\ket{j}$ and~$\ket{k}$. For rotations around the~$x-$ axis we use the following notations:~$\hat{R}_{x}^{jk}(\theta) := \hat{R}_{\phi = 0}^{jk}(\theta),\ \hat{R}_{-x}^{jk}(\theta) := \hat{R}_{\phi = \pi}^{jk}(\theta)$. The second type of operation is the generalized qudit rotation around the~$z-$ axis. This is simply a phase shift applied to the state~$\ket j$:
\begin{equation}
    \hat{R}_z^{j}(\theta) = \exp( i \theta \ketbra{j}{j}).
\end{equation}
\begin{figure}[t]
\centering
\subfloat[\label{fig:qudit-spin-echo_zero_ringbauer}]{
    \includegraphics[height = 0.40\linewidth]{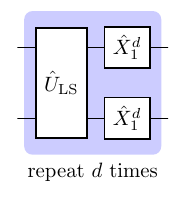}
}
\subfloat[\label{fig:qudit-spin-echo_zero_optimal}]{
    \includegraphics[height = 0.40\linewidth]{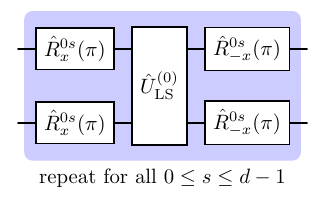}
}
\hfill
\subfloat[\label{fig:qudit-spin-echo_modified}]{
    \includegraphics[width = 0.70\linewidth]{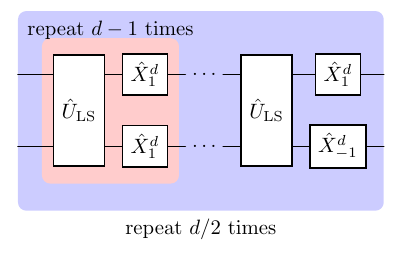}
}
\caption{Spin-echo sequences for the LS
gate. (a) The spin-echo sequence from Ref. \cite{hrmo2023Native}. (b) The spin-echo sequence that is valid for the zero-order LS gate. (c) The spin-echo sequence that can be used for a general LS gate with an even qudit dimension~$d$.}
\label{fig:spin echo sequences}
\end{figure}
\subsection{Type (a) sequence, zero-order case}
We start our investigation from the spin-echo sequence proposed previously in Ref.~\cite{hrmo2023Native}, which we call a  type~(a) sequence. It is illustrated in Fig.~\ref{fig:qudit-spin-echo_zero_ringbauer}. The sequence consists of~$d$ loops, where each loop has the form~$(\hat{X}_1^d \otimes \hat{X}_1^d ) \cdot \hat{U}_{\mathrm{LS}} $. 
The operator~$\hat{X}_m^d$ performs a cyclic permutation by~$m$:
\begin{equation}
    \hat{X}_m^d\ket{j}\propto\ket{(j+m)\ \mathrm{mod} \ d},
\end{equation}
and can be realized using~$O(d)$ native single-qudit rotations (See Appendix~\ref{sec:appendix_X_m}).
For convenience, we define the following subspaces:
\begin{equation}
    \label{eq:subspace G_l}
    \mathbf{G}_{l}^d = \mathrm{span}\left(\left\{  \ket{ss'}  \vert (s -s')\ \mathrm{mod} \ d = l \right\}\right).
\end{equation}

\begin{figure}[t]
\centering
\includegraphics[width=6.285 cm]{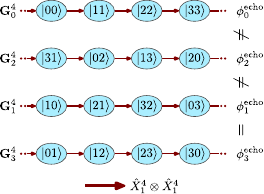}
\caption{A scheme that shows how different states acquire the phase during the spin-echo sequence of type (a) for ququarts ($d = 4$).}
\label{fig:spin-echo-demonstration-ringbauer}
\end{figure}

Note that the operator~$\hat{X}_m^d \otimes \hat{X}_m^d$ acting on~$\ket{ss'}$ preserves the quantity~$(s -s')\ \mathrm{mod} \ d$. Therefore, the subspaces are invariant with respect to the spin-echo operators. Moreover, since each LS gate loop is identical, every logical state $\ket{ss'} \in \mathbf{G}_{l}^d$ spends an equal amount of time in each other logical state of its subspace, including itself (see Fig.~\ref{fig:spin-echo-demonstration-ringbauer}). Thus, states within the same subspace acquire identical phases:
\begin{equation}
    \label{eq:ringbauer_spinecho}
    \phi^{(a)}_{ss'}  = \phi_l^{\mathrm{echo}} ,  \ \mathrm{if} \ \ket{ss'} \in \mathbf{G}_{l}^d, 
\end{equation}
where~$\phi^{(a)}_{ss'}$ is the final phase acquired by the state~$\ket{ss'}$ after the sequence of type (a). The phase~$\phi_l^{\mathrm{echo}}$:
\begin{equation}
\label{eq:LS_spin_echo_phi_l}
    \phi_l^{\mathrm{echo}} = \sum \limits_{s}\phi^{\mathrm{LS}}_{s,\,(s+l)\ \mathrm{mod} \ d}.
\end{equation}
Using the symmetry relation~$\phi^{\mathrm{LS}}_{ss'} = \phi^{\mathrm{LS}}_{s's}$, we find that ~$\phi_{d-l}^{\mathrm{echo}} = \phi_l^{\mathrm{echo}}$ for all~$l>0$. Hence, after the spin-echo sequence of type (a) we obtain~$\lfloor d/2\rfloor + 1$ distinct LS phases in the general case. Therefore, for~$d>3$ the sequence of type (a) does not provide the maximum reduction of the LS gate (see Fig. \ref{fig:spin-echo-demonstration-ringbauer}). However, in the zero-order case~\eqref{eq:evolution_LS_0}, which took place in Ref.~\cite{hrmo2023Native},
only~$\phi_{00}^{\mathrm{LS}}$ is non-zero and other phases are negligible:
\begin{equation}
\label{eq:LS_spin-echo_zero_ord}
    \phi_{ss'} = \begin{cases}
        \phi_{00}^{\mathrm{LS}}, \quad &\mathrm{if} \ s = s'=0, \\
        0, \quad &\mathrm{otherwise}.
    \end{cases}
\end{equation}
Using Eq.~\eqref{eq:LS_spin_echo_phi_l}:
\begin{equation}
    \phi_{l}^{\mathrm{echo}} = \begin{cases}
        \phi_{00}^{\mathrm{LS}}, \quad &\mathrm{if} \ l = 0, \\
        0, \quad &\mathrm{otherwise}.
    \end{cases}
\end{equation}
Defining~$\mathbf{H}_0 = \mathbf{G}_0,\ \mathbf{H}_1 = \bigoplus_{l\neq 0} \mathbf{G}_l$, we achieve the desired LS gate reduction.
\subsection{\textcolor{black}{Type (b) sequence, zero-order case}}
\label{sec:spin_echo_b}
The sequence of type (a) requires $d(d-1) = O(d^2)$ native single-qudit gates: $(d-1)$ gates for each cyclic shift times~$d$ spin-echo pulses. However, in the zero-order case, to which this scheme is anyway only applicable to, this quadratic scaling with~$d$ can be reduced using a sequence of type (b), which we propose here. The sequence has the following operator representation:
\begin{equation}
\label{eq:LS_spin-echo zero optimal}
    \hat{U}_{\mathrm{LS}}^{(b)} = \prod_{s = 0}^{d-1} \hat{R}_{-x}^{0s}(\pi)^{\otimes 2} \hat{U}_{\mathrm{LS}}^{(0)} \hat{R}_{x}^{0s}(\pi)^{\otimes 2},
\end{equation}
where~$\hat{R}_{x}^{00}(\pi) = \hat{\mathbb{1}}$.
The sequence is shown in Fig.~\ref{fig:qudit-spin-echo_zero_optimal}. It also consists of~$d$ cycles, so the number of LS gates is the same. In the~$s^{\mathrm{th}}$ loop we swap the population between states~$\ket{00}$ and $\ket{ss}$, ensuring each~$\ket{ss}$ acquires the same phase~$\phi_{00}^{\mathrm{LS}}$, while leaving the phases of other states with~$s\neq s'$ unaltered. After the LS gate action we reverse the swap around the~$-x$ axis, so that no additional phases accumulate. The LS phases after this sequence are similar to the phases after the spin-echo sequence of type (a), i.e.~$\phi^{(b)} = \phi^{(a)}$. However, the number of native single-qudit rotations reduces to~$2(d-1) = O(d)$. 

\begin{figure}[t]
\centering
\includegraphics[width=8 cm]{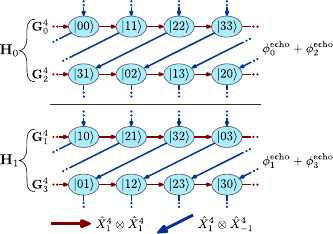}
\caption{A schematic representation of the phase accumulation for the logical states of two qudits during the spin-echo sequence of type (c), applied to ququarts ($d = 4$).}
\label{fig:spin-echo-demonstration-modified}
\end{figure}

\subsection{Type (c) sequence, general case for even $d$}
\label{sec:spin_echo_general}
In the following, we propose a spin-echo sequence that can maximally reduce the LS gate in the general case for even~$d$. First, we separate the subspaces~$\mathbf{G}_{l}^d $ into two different groups with even and odd indices~$l$:
\begin{equation}
    \mathbf{H}_{0} = \bigoplus_{l -\mathrm{even}} \mathbf{G}_{l}^d, \quad \mathbf{H}_{1} = \bigoplus_{l -\mathrm{odd}} \mathbf{G}_{l}^d.
\end{equation}
The proposed spin-echo sequence, which we refer to as the type (c) sequence, consists of~$d/2$ cycles, where each cycle has the form 
\begin{equation}
\label{eq:loop_modified spin-echo}
\left( \hat{X}_{1}^d \otimes \hat{X}_{-1}^d \right) \hat{U}_{\mathrm{LS}} \left( \left( \hat{X}_{1}^d \otimes \hat{X}_{1}^d \right) \hat{U}_{\mathrm{LS}} \right)^{d-1}.
\end{equation}
The circuit is shown in Fig.~\ref{fig:qudit-spin-echo_modified}. To understand the principles behind this scheme, it is instructive to study the example presented in Fig.~\ref{fig:spin-echo-demonstration-modified} for~$d=4$. The first~$d-1$ operators~$\hat{X}_{1}^d \otimes \hat{X}_{1}^d$ of the cycle shift each physical state 'along' their respective subspaces~$\mathbf{G}_{l}^d$, since they are invariant. Then the last operator~$ \hat{X}_{1}^d \otimes \hat{X}_{-1}^d$ of the cycle shifts each~$\mathbf{G}_{l}^d$ 'downwards', so that~$\mathbf{G}_{l}^d \rightarrow \mathbf{G}_{(l + 2)\, \mathrm{mod} \, d}^d$. In the end, every basis state of~$\mathbf{H}_{0}$ or~$\mathbf{H}_{1}$ will be visited exactly once, independent of the starting state. Therefore, the LS phase for an arbitrary state~$\ket{\psi}$ in one of the subspaces will amount to one of two values:
\begin{equation}
    \label{eq:phi_c}
    \phi^{(c)}(\psi)  = \begin{cases}
        \sum\limits_{l -\mathrm{even}} \phi_l^{\mathrm{echo}} ,  \  \ &\ket{\psi} \in \mathbf{H}_{0}, \\
        \sum\limits_{l -\mathrm{odd}}\phi_l^{\mathrm{echo}} ,  \  &\ket{\psi} \in \mathbf{H}_{1}, \\
    \end{cases}
\end{equation}
which achieves the desired reduction. Note that instead of the operator~$ \hat{X}_{1}^d \otimes \hat{X}_{-1}^d$ a variety of other operators can be used at the end of the cycle to shift the subspaces~$\mathbf{G}_{l}^d$. That is, any operator of the form~$\hat{X}_{(n+2)\ \mathrm{mod}\ d}^{d} \otimes \hat{X}_n^d$. We select~$ \hat{X}_{1}^d \otimes \hat{X}_{-1}^d$, since it helps balance the number of single-qudit rotations between the two qudits, thus suppressing the accumulation of errors. Additionally, a cyclic shift with a unit step is straightforward to execute, unlike any other step (see Appendix~\ref{sec:appendix_X_m}). Note that native rotations comprising the shifts can occasionally be performed around the~$-x$ axis, instead of~$+x$, in order to compensate for over-rotation, as was done in Ref.~\cite{hrmo2023Native}. The total number of the two-particle gates equals~$d^2/2$, and the number of single qudit gates is~$d^3/2$. 

\begin{figure}
\resizebox{0.40\linewidth}{!}{
\Large
    \centering
    \includegraphics[height = 0.80\linewidth]{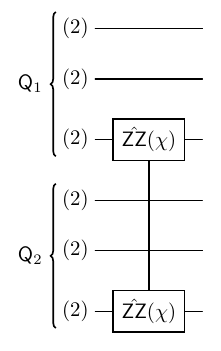}
}
    \caption{Operation~$\hat{\mathsf{ZZ}}(\chi) := \exp(i \chi \hat{\sigma}_Z \otimes \hat{\sigma}_Z)$ on the last qubits embedded in different quocts($d = 8$), which is equivalent to the type (c) spin-echo sequence.}
    \label{fig:ZZ_embedded}
\end{figure}
It can be shown (see Appendix~\ref{sec:spin-echo-embedded}) that the proposed sequence directly implements a~$\hat{\sigma}_Z \otimes \hat{\sigma}_Z$ gate between two embedded qubits in different qudits, as long as~$d$ is a power of 2 (Fig.~\ref{fig:ZZ_embedded}). The term "directly" means that the implementation does not require any additional single-particle gates.

As a final note, we point out that the sequence of type (c) only works for even~$d$. The reason is that, for odd~$d$, the spaces~$\mathbf{H}_{i}$ are not invariant under~$\hat{X}_{(n+2)\ \mathrm{mod}\ d}^{d} \otimes \hat{X}_n^d$, which maps the spaces~$\mathbf{G}_{l}^d$ onto each other. However, a partial LS gate reduction can be achieved with~$\hat{X}_{(n+p)\ \mathrm{mod}\ d}^{d} \otimes \hat{X}_n^d$, where~$p$ is the smallest prime divisor of~$d$. The spaces~$\mathbf{G}_{l}^d$ should then be gathered into~$\mathbf{H}_{i}$ by their value of~$l$ modulo~$p$. By that construction, there will be~$p$ distinct phases.

\subsection{Local phase compensation with spin echo}
\label{sec:
non_entangling_echo}
While we previously considered only the nonlocal term of the LS gate, we now examine the local term~\eqref{eq:evolution_LS_single}. It is evident that the phases $\chi_{jj;ss'}^{\mathrm{LS}}$ accumulated in sequences of types (a) and (c) are all equal, since in these sequences each ion spends an equal amount of time in each of its~$d$ basis states. Therefore, these phases can be omitted as global. To see that the type (b) sequence also has this property in the zero-order approximation, we note that, for laser-induced Stark shifts, the following inequality holds:~$| \delta_{j,s}^{\mathrm{AC}} | \leq |\Omega_{j,s}|$. Hence, in the zero-order case, for which the sequence was designed, one can ignore all terms in Eq.~\eqref{eq:Phi_j LS} except the zero-state term:
\begin{equation}
    \hat{\Phi}_{j}^{(0)} \approx 
\left( \chi_{jj}^\mathrm{LS} + \phi_{j,0}^\mathrm{LS} \right) \ketbra{0_j}{0_j}.
\end{equation}
Since only the~$\ket{0}$ state acquires a non-negligible phase, and each ion spends an equal time in this state during the sequence, the local phases can be omitted, similarly to the cases described above. Therefore, spin-echo sequences of all types automatically provide local phase cancellation.
\section{Discussion and conclusion}
\label{sec:discussion}
In this paper we analyzed generalizations of MS and LS gates on trapped-ion qudits. We considered the most general case of these gates with arbitrary pulse shaping and taking into account multiple motional modes. We have shown that taking several modes into account can, for example, lead to different local phases acquired by two ions during the MS gate and studied the general form of the qudit LS gate. We also compared the robustness of these gates to fluctuations of the experimental parameters between them and with their qubit counterparts. 

The simplest structure that facilitates efficient transpilation of the circuits for qubits embedded into qudits architecture and robustness is expressed by the MS and zero-order LS gates. That makes them the most suitable options for a universal qudit processor implementation. The MS gate due to its resonant nature usually requires less optical power, however is more sensitive to the laser spectral properties and is harder to use for mixed-species gates~\cite{sawyer2021Wavelengthinsensitive, hughes2020Benchmarking, ballance2015Hybrid, inlek2017Multispecies, tan2015Multielement, negnevitsky2018Repeated, bruzewicz2019Dualspecies, wan2019Quantum}.
Moreover, the reduced number of controlled phases leads to higher fidelity of two-particle gates. For instance, the MS gate can be effectively used for implementation of a controlled-Z ($\mathsf{CZ}$) gate between two embedded qubits within a pair of ququarts using a single two-particle operation \cite{nikolaeva2024Universal} by setting the entangling phase~$\chi_{12}^{\mathrm{MS}}= \pi/2$. Similarly, a zero-order LS gate with~$\chi_{12}^{\mathrm{LS}}= \pi/2$ allows to implement a four-qubit~$\mathsf{CZ}$-gate, creating multi-qubit entanglement in a single two-particle gate~\cite{nikolaeva2024Efficient}, realizing multi-qubit entanglement. The more general forms of the LS gate, however, may offer greater utility for the quantum simulation of natively high-dimensional quantum systems.

We also analyze the overhead in number of local and nonlocal phases in qudits compared to qubits. A qudit MS gate has one nonlocal phase
and two local phases, usually neglected as global for the qubit counterparts. A general LS gate has $d(d-1)/2$ nonlocal and $d-1$ local phases, presenting a significantly larger set of parameters to control. The zero-order LS-gate, however, has only one local and one nonlocal phase each. Despite this overhead in control parameters, we demonstrate 
that the robustness of the MS gate can be significantly improved using the same pulse shaping techniques that are well established for qubit processors. Moreover, we have shown that the pulse shapes optimized for the stability of the nonlocal phase to secular frequency drifts provide almost the same stability for the local phases. The reason for this is the interdependence of local and nonlocal phases, meaning that stabilizing one of the phases implicitly increases the robustness of others.

We also studied the methods for simplifying the structure of the LS gate using the spin-echo method, that helps to reduce the overhead in control parameters with respect to qubits. In the resulting gate, the number of nonlocal phases is reduced to one, while the local phase become compensated. 
\textcolor{black}{The spin-echo sequence of type (b) proposed in section~\ref{sec:spin_echo_b} makes use of the zero-order approximation, which turns out to be a requirement for the type (a) sequence to function properly. This approximation allows for the reduced scaling of the number of native single-qudit gates from~$O(d^2)$ for type (a) to~$O(d)$ for type (b).}

The spin-echo sequence of type (c) proposed in section~\ref{sec:spin_echo_general} has the~$O(d^2)$ scaling on the required number of LS gates. \textcolor{black}{Despite this method looking considerably less appealing than the type (a) method, which only requires~$O(d)$ two-qudit gates}, we believe that, in the most general case, no spin-echo sequence that reduces the number of distinct two-qudit phases to two (or any fixed number independent of~$d$) can have better scaling. Our reasoning is based on the following. There are~$O(d^2)$ LS phases~$\phi_{ss'}$, and gate reduction ultimately comes down to distributing these phases among a fixed number of "boxes", with the sum of all phases in each box being the total accumulated spin-echo phase. The number of phases in each box grows quadratically with~$d$, since they contain a fixed fraction of the entire set of phases. To perform the sequence, one has to move a specific basis state across the states in a box while performing LS gates on each of those intermediate states. Thus, the number of LS gates should also scale quadratically. This argument is further reinforced by the result obtained in Ref.~\cite{nikolaeva2024Efficient}, where the number of two-particle gates needed to realize a controlled-Z gate on embedded qubits also grows quadratically with the qudit dimension. A single-qudit cyclic shift (or any permutation in general) requires~$O(d)$ native gates, therefore, the scaling for the native gate number cannot be better than~$d$ times the scaling for the number of LS gates, that is, $O(d^3)$. The only exception is the zero-order case, for which there are~$d$ non-zero LS phases, and so the type (a) and (b) sequences have the~$O(d)$ scaling. Therefore, if possible, one should try to preemptively eliminate as many phases as possible by choosing a more convenient qudit space, suitable polarizations, etc. to avoid the need of too many two-particle gates.

In general, we believe that the methods studied in this work should be applicable in a variety of different cases, depending on the physical properties of the ion and the technical limitations of a specific experimental setup. When using qudits, one should pay close attention to the qudit phases and make corresponding changes to the conventional two-qubit entangling gate design.
\section*{Acknowledgments}
This research was supported by the Russian State Corporation "Rosatom" within the Roadmap on Quantum Computing, Contract No. 868/1653-D dated 21 August 2025. Additionally, we would like to thank Anastasiia Nikolaeva and Evgeniy Kiktenko for helpful discussions and important contributions regarding the local equivalence of two-qudit gates. 
\bibliography{ref_final}

@article{aksenov2023Realizing,
  title = {Realizing Quantum Gates with Optically Addressable {{Yb}} + 171 Ion Qudits},
  author = {Aksenov, M. A. and Zalivako, I. V. and Semerikov, I. A. and Borisenko, A. S. and Semenin, N. V. and Sidorov, P. L. and Fedorov, A. K. and Khabarova, K. {\relax Yu}. and Kolachevsky, N. N.},
  year = 2023,
  month = may,
  journal = {Physical Review A},
  volume = {107},
  number = {5},
  pages = {052612},
  issn = {2469-9926, 2469-9934},
  doi = {10.1103/PhysRevA.107.052612},
  urldate = {2024-07-23},
  langid = {english}
}

@article{aolita2007Highfidelity,
  title = {High-Fidelity Ion-Trap Quantum Computing with Hyperfine Clock States},
  author = {Aolita, L. and Kim, K. and Benhelm, J. and Roos, C. F. and H{\"a}ffner, H.},
  year = 2007,
  month = oct,
  journal = {Physical Review A},
  volume = {76},
  number = {4},
  pages = {040303},
  issn = {1050-2947, 1094-1622},
  doi = {10.1103/PhysRevA.76.040303},
  urldate = {2024-07-23},
  copyright = {http://link.aps.org/licenses/aps-default-license},
  langid = {english}
}

@article{arute2019Quantum,
  title = {Quantum Supremacy Using a Programmable Superconducting Processor},
  author = {Arute, Frank and Arya, Kunal and Babbush, Ryan and Bacon, Dave and Bardin, Joseph C. and Barends, Rami and Biswas, Rupak and Boixo, Sergio and Brandao, Fernando GSL and Buell, David A.},
  year = 2019,
  journal = {Nature},
  volume = {574},
  number = {7779},
  pages = {505--510},
  publisher = {Nature Publishing Group},
  urldate = {2024-07-27}
}

@article{baldwin2021Highfidelity,
  title = {High-Fidelity Light-Shift Gate for Clock-State Qubits},
  author = {Baldwin, C. H. and Bjork, B. J. and {Foss-Feig}, M. and Gaebler, J. P. and Hayes, D. and Kokish, M. G. and Langer, C. and Sedlacek, J. A. and Stack, D. and Vittorini, G.},
  year = 2021,
  month = jan,
  journal = {Physical Review A},
  volume = {103},
  number = {1},
  pages = {012603},
  issn = {2469-9926, 2469-9934},
  doi = {10.1103/PhysRevA.103.012603},
  urldate = {2024-07-23},
  langid = {english}
}

@article{ballance2015Hybrid,
  title = {Hybrid Quantum Logic and a Test of {{Bell}}'s Inequality Using Two Different Atomic Isotopes},
  author = {Ballance, C. J. and Sch{\"a}fer, V. M. and Home, Jonathan P. and Szwer, D. J. and Webster, Scott C. and Allcock, D. T. C. and Linke, Norbert M. and Harty, T. P. and Aude Craik, D. P. L. and Stacey, Derek N.},
  year = 2015,
  journal = {Nature},
  volume = {528},
  number = {7582},
  pages = {384--386},
  publisher = {Nature Publishing Group UK London},
  urldate = {2025-10-08}
}

@article{bian2026Architecture,
  title = {Architecture of a Scalable Universal Quantum Processor by Encoding Two Qubits on Electron and Nuclear Spins in a Trapped Ion},
  author = {Bian, Ji and Liu, Teng and Ding, Min and Lao, Qifeng and Zhang, Huiyi and Rao, Xinxin and Lu, Pengfei and Luo, Le},
  year = 2026,
  journal = {New Journal of Physics},
  volume = {28},
  number = {1},
  pages = {014509},
  publisher = {IOP Publishing},
  urldate = {2026-03-05}
}

@article{blanes2009Magnus,
  title = {The {{Magnus}} Expansion and Some of Its Applications},
  author = {Blanes, Sergio and Casas, Fernando and Oteo, Jose-Angel and Ros, Jos{\'e}},
  year = 2009,
  journal = {Physics reports},
  volume = {470},
  number = {5-6},
  pages = {151--238},
  publisher = {Elsevier},
  urldate = {2024-07-26}
}

@article{blumel2021Poweroptimal,
  title = {Power-Optimal, Stabilized Entangling Gate between Trapped-Ion Qubits},
  author = {Bl{\"u}mel, Reinhold and Grzesiak, Nikodem and Pisenti, Neal and Wright, Kenneth and Nam, Yunseong},
  year = 2021,
  journal = {npj Quantum Information},
  volume = {7},
  number = {1},
  pages = {147},
  publisher = {Nature Publishing Group UK London},
  urldate = {2024-07-23}
}

@article{bluvstein2024Logical,
  title = {Logical Quantum Processor Based on Reconfigurable Atom Arrays},
  author = {Bluvstein, Dolev and Evered, Simon J. and Geim, Alexandra A. and Li, Sophie H. and Zhou, Hengyun and Manovitz, Tom and Ebadi, Sepehr and Cain, Madelyn and Kalinowski, Marcin and Hangleiter, Dominik},
  year = 2024,
  journal = {Nature},
  volume = {626},
  number = {7997},
  pages = {58--65},
  publisher = {Nature Publishing Group UK London},
  urldate = {2024-07-27}
}

@misc{brennen2005Efficient,
  title = {Efficient {{Circuits}} for {{Exact-Universal Computations}} with {{Qudits}}},
  author = {Brennen, Gavin K. and Bullock, Stephen S. and O'Leary, Dianne P.},
  year = 2005,
  month = sep,
  number = {arXiv:quant-ph/0509161},
  eprint = {quant-ph/0509161},
  publisher = {arXiv},
  doi = {10.48550/arXiv.quant-ph/0509161},
  urldate = {2025-09-24},
  archiveprefix = {arXiv}
}

@article{bruzewicz2019Dualspecies,
  title = {Dual-Species, Multi-Qubit Logic Primitives for {{Ca}}+/{{Sr}}+ Trapped-Ion Crystals},
  author = {Bruzewicz, C. D. and McConnell, R. and Stuart, J. and Sage, J. M. and Chiaverini, J.},
  year = 2019,
  journal = {npj Quantum Information},
  volume = {5},
  number = {1},
  pages = {102},
  publisher = {Nature Publishing Group UK London},
  urldate = {2025-10-08}
}

@article{bruzewicz2019Trappedion,
  title = {Trapped-Ion Quantum Computing: {{Progress}} and Challenges},
  shorttitle = {Trapped-Ion Quantum Computing},
  author = {Bruzewicz, Colin D. and Chiaverini, John and McConnell, Robert and Sage, Jeremy M.},
  year = 2019,
  journal = {Applied Physics Reviews},
  volume = {6},
  number = {2},
  publisher = {AIP Publishing},
  urldate = {2024-07-24}
}

@article{cheng2024Crosstalk,
  title = {Crosstalk Suppression of Parallel Gates for Fault-Tolerant Quantum Computation with Trapped Ions via Optical Tweezers},
  author = {Cheng, Lin and Liu, Sheng-Chen and Peng, Liang-You and Gong, Qihuang},
  year = 2024,
  month = sep,
  journal = {Physical Review Applied},
  volume = {22},
  number = {3},
  pages = {034021},
  issn = {2331-7019},
  doi = {10.1103/PhysRevApplied.22.034021},
  urldate = {2025-01-20},
  langid = {english}
}

@article{choi2014Optimal,
  title = {Optimal {{Quantum Control}} of {{Multimode Couplings}} between {{Trapped Ion Qubits}} for {{Scalable Entanglement}}},
  author = {Choi, T. and Debnath, S. and Manning, T. A. and Figgatt, C. and Gong, Z.-X. and Duan, L.-M. and Monroe, C.},
  year = 2014,
  month = may,
  journal = {Physical Review Letters},
  volume = {112},
  number = {19},
  pages = {190502},
  issn = {0031-9007, 1079-7114},
  doi = {10.1103/PhysRevLett.112.190502},
  urldate = {2024-07-23},
  copyright = {http://link.aps.org/licenses/aps-default-license},
  langid = {english}
}

@article{cirac1995Quantum,
  title = {Quantum {{Computations}} with {{Cold Trapped Ions}}},
  author = {Cirac, J. I. and Zoller, P.},
  year = 1995,
  month = may,
  journal = {Physical Review Letters},
  volume = {74},
  number = {20},
  pages = {4091--4094},
  issn = {0031-9007, 1079-7114},
  doi = {10.1103/PhysRevLett.74.4091},
  urldate = {2024-07-23},
  copyright = {http://link.aps.org/licenses/aps-default-license},
  langid = {english}
}

@article{debnath2016Demonstration,
  title = {Demonstration of a Small Programmable Quantum Computer with Atomic Qubits},
  author = {Debnath, Shantanu and Linke, Norbert M. and Figgatt, Caroline and Landsman, Kevin A. and Wright, Kevin and Monroe, Christopher},
  year = 2016,
  journal = {Nature},
  volume = {536},
  number = {7614},
  pages = {63--66},
  publisher = {Nature Publishing Group UK London},
  urldate = {2024-07-23}
}

@article{ebadi2021Quantum,
  title = {Quantum Phases of Matter on a 256-Atom Programmable Quantum Simulator},
  author = {Ebadi, Sepehr and Wang, Tout T. and Levine, Harry and Keesling, Alexander and Semeghini, Giulia and Omran, Ahmed and Bluvstein, Dolev and Samajdar, Rhine and Pichler, Hannes and Ho, Wen Wei},
  year = 2021,
  journal = {Nature},
  volume = {595},
  number = {7866},
  pages = {227--232},
  publisher = {Nature Publishing Group UK London},
  urldate = {2024-07-27}
}

@article{fedorov2012Implementation,
  title = {Implementation of a {{Toffoli}} Gate with Superconducting Circuits},
  author = {Fedorov, Arkady and Steffen, Lars and Baur, Matthias and {da Silva}, Marcus P. and Wallraff, Andreas},
  year = 2012,
  journal = {Nature},
  volume = {481},
  number = {7380},
  pages = {170--172},
  publisher = {Nature Publishing Group UK London},
  urldate = {2024-07-27}
}

@misc{fedorov2022Quantum,
  title = {Quantum Computing at the Quantum Advantage Threshold: A down-to-Business Review},
  shorttitle = {Quantum Computing at the Quantum Advantage Threshold},
  author = {Fedorov, A. K. and Gisin, N. and Beloussov, S. M. and Lvovsky, A. I.},
  year = 2022,
  month = mar,
  number = {arXiv:2203.17181},
  eprint = {2203.17181},
  primaryclass = {physics, physics:quant-ph},
  publisher = {arXiv},
  urldate = {2024-07-24},
  archiveprefix = {arXiv}
}

@article{gerster2022Experimental,
  title = {Experimental {{Bayesian Calibration}} of {{Trapped-Ion Entangling Operations}}},
  author = {Gerster, Lukas and {Mart{\'i}nez-Garc{\'i}a}, Fernando and Hrmo, Pavel and Van Mourik, Martin W. and Wilhelm, Benjamin and Vodola, Davide and M{\"u}ller, Markus and Blatt, Rainer and Schindler, Philipp and Monz, Thomas},
  year = 2022,
  month = jun,
  journal = {PRX Quantum},
  volume = {3},
  number = {2},
  pages = {020350},
  issn = {2691-3399},
  doi = {10.1103/PRXQuantum.3.020350},
  urldate = {2024-07-29},
  langid = {english}
}

@inproceedings{grover1996fast,
  title = {A Fast Quantum Mechanical Algorithm for Database Search},
  booktitle = {Proceedings of the Twenty-Eighth Annual {{ACM}} Symposium on {{Theory}} of Computing  - {{STOC}} '96},
  author = {Grover, Lov K.},
  year = 1996,
  pages = {212--219},
  publisher = {ACM Press},
  address = {Philadelphia, Pennsylvania, United States},
  doi = {10.1145/237814.237866},
  urldate = {2024-07-27},
  isbn = {978-0-89791-785-8},
  langid = {english}
}

@article{guo2024siteresolved,
  title = {A Site-Resolved Two-Dimensional Quantum Simulator with Hundreds of Trapped Ions},
  author = {Guo, S.-A. and Wu, Y.-K. and Ye, J. and Zhang, L. and Lian, W.-Q. and Yao, R. and Wang, Y. and Yan, R.-Y. and Yi, Y.-J. and Xu, Y.-L.},
  year = 2024,
  journal = {Nature},
  pages = {1--6},
  publisher = {Nature Publishing Group UK London},
  urldate = {2024-11-09}
}

@article{haddadfarshi2016High,
  title = {High Fidelity Quantum Gates of Trapped Ions in the Presence of Motional Heating},
  author = {Haddadfarshi, Farhang and Mintert, Florian},
  year = 2016,
  journal = {New Journal of Physics},
  volume = {18},
  number = {12},
  pages = {123007},
  publisher = {IOP Publishing},
  urldate = {2024-07-26}
}

@article{harrow2009Quantum,
  title = {Quantum {{Algorithm}} for {{Linear Systems}} of {{Equations}}},
  author = {Harrow, Aram W. and Hassidim, Avinatan and Lloyd, Seth},
  year = 2009,
  month = oct,
  journal = {Physical Review Letters},
  volume = {103},
  number = {15},
  pages = {150502},
  issn = {0031-9007, 1079-7114},
  doi = {10.1103/PhysRevLett.103.150502},
  urldate = {2024-07-27},
  copyright = {http://link.aps.org/licenses/aps-default-license},
  langid = {english}
}

@article{henriet2020Quantum,
  title = {Quantum Computing with Neutral Atoms},
  author = {Henriet, Lo{\"i}c and Beguin, Lucas and Signoles, Adrien and Lahaye, Thierry and Browaeys, Antoine and Reymond, Georges-Olivier and Jurczak, Christophe},
  year = 2020,
  journal = {Quantum},
  volume = {4},
  pages = {327},
  publisher = {Verein zur F\"orderung des Open Access Publizierens in den Quantenwissenschaften},
  urldate = {2024-07-27}
}

@misc{hill2021Realization,
  title = {Realization of Arbitrary Doubly-Controlled Quantum Phase Gates},
  author = {Hill, Alexander D. and Hodson, Mark J. and Didier, Nicolas and Reagor, Matthew J.},
  year = 2021,
  month = aug,
  number = {arXiv:2108.01652},
  eprint = {2108.01652},
  primaryclass = {quant-ph},
  publisher = {arXiv},
  urldate = {2024-07-27},
  archiveprefix = {arXiv}
}

@article{hrmo2023Native,
  title = {Native Qudit Entanglement in a Trapped Ion Quantum Processor},
  author = {Hrmo, Pavel and Wilhelm, Benjamin and Gerster, Lukas and {van Mourik}, Martin W. and Huber, Marcus and Blatt, Rainer and Schindler, Philipp and Monz, Thomas and Ringbauer, Martin},
  year = 2023,
  journal = {Nature Communications},
  volume = {14},
  number = {1},
  pages = {2242},
  publisher = {Nature Publishing Group UK London},
  urldate = {2024-09-07}
}

@article{hughes2020Benchmarking,
  title = {Benchmarking a {{High-Fidelity Mixed-Species Entangling Gate}}},
  author = {Hughes, A. C. and Sch{\"a}fer, V. M. and Thirumalai, K. and Nadlinger, D. P. and Woodrow, S. R. and Lucas, D. M. and Ballance, C. J.},
  year = 2020,
  month = aug,
  journal = {Physical Review Letters},
  volume = {125},
  number = {8},
  pages = {080504},
  issn = {0031-9007, 1079-7114},
  doi = {10.1103/PhysRevLett.125.080504},
  urldate = {2025-10-08},
  langid = {english}
}

@misc{hughes2025Trappedion,
  title = {Trapped-Ion Two-Qubit Gates with {$>$}99.99\% Fidelity without Ground-State Cooling},
  author = {Hughes, A. C. and Srinivas, R. and L{\"o}schnauer, C. M. and Knaack, H. M. and Matt, R. and Ballance, C. J. and Malinowski, M. and Harty, T. P. and Sutherland, R. T.},
  year = 2025,
  month = oct,
  number = {arXiv:2510.17286},
  eprint = {2510.17286},
  primaryclass = {quant-ph},
  publisher = {arXiv},
  doi = {10.48550/arXiv.2510.17286},
  urldate = {2025-10-25},
  archiveprefix = {arXiv}
}

@article{franke2023quantum,
  title={Quantum-enhanced sensing on optical transitions through finite-range interactions},
  author={Franke, Johannes and Muleady, Sean R and Kaubruegger, Raphael and Kranzl, Florian and Blatt, Rainer and Rey, Ana Maria and Joshi, Manoj K and Roos, Christian F},
  journal={Nature},
  volume={621},
  number={7980},
  pages={740--745},
  year={2023},
  publisher={Nature Publishing Group UK London}
}

@article{kirchmair2009deterministic,
  title={Deterministic entanglement of ions in thermal states of motion},
  author={Kirchmair, G and Benhelm, J and Z{\"a}hringer, F and Gerritsma, R and Roos, Christian F and Blatt, Rainer},
  journal={New Journal of Physics},
  volume={11},
  number={2},
  pages={023002},
  year={2009},
  publisher={IOP Publishing}
}

@article{semenin2026equivalence,
  title={On the local equivalence of trapped-ion two-qudit gates},
  author={Semenin, Nikita V and Kamenskikh, Pavel A and Zalivako, Ilia V and Nikolaeva, Anastasiia S and Kiktenko, Evgeniy O},
  journal={arXiv preprint arXiv:2606.03643},
  url={https://arxiv.org/abs/2606.03643},
  year={2026}
}

@article{inlek2017Multispecies,
  title = {Multispecies {{Trapped-Ion Node}} for {{Quantum Networking}}},
  author = {Inlek, I. V. and Crocker, C. and Lichtman, M. and Sosnova, K. and Monroe, C.},
  year = 2017,
  month = jun,
  journal = {Physical Review Letters},
  volume = {118},
  number = {25},
  pages = {250502},
  issn = {0031-9007, 1079-7114},
  doi = {10.1103/PhysRevLett.118.250502},
  urldate = {2025-10-08},
  copyright = {http://link.aps.org/licenses/aps-default-license},
  langid = {english}
}

@techreport{james1998Quantum,
  title = {Quantum Dynamics of Cold Trapped Ions with Application to Quantum Computation},
  author = {James, Daniel FV},
  year = 1998,
  urldate = {2024-07-23}
}

@article{kazmina2024Demonstration,
  title = {Demonstration of a Parity-Time-Symmetry-Breaking Phase Transition Using Superconducting and Trapped-Ion Qutrits},
  author = {Kazmina, Alena S. and Zalivako, Ilia V. and Borisenko, Alexander S. and Nemkov, Nikita A. and Nikolaeva, Anastasiia S. and Simakov, Ilya A. and Kuznetsova, Arina V. and Egorova, Elena Yu. and Galstyan, Kristina P. and Semenin, Nikita V. and Korolkov, Andrey E. and Moskalenko, Ilya N. and Abramov, Nikolay N. and Besedin, Ilya S. and Kalacheva, Daria A. and Lubsanov, Viktor B. and Bolgar, Aleksey N. and Kiktenko, Evgeniy O. and Khabarova, Ksenia Yu. and Galda, Alexey and Semerikov, Ilya A. and Kolachevsky, Nikolay N. and Maleeva, Nataliya and Fedorov, Aleksey K.},
  year = 2024,
  month = mar,
  journal = {Physical Review A},
  volume = {109},
  number = {3},
  pages = {032619},
  issn = {2469-9926, 2469-9934},
  doi = {10.1103/PhysRevA.109.032619},
  urldate = {2024-07-23},
  langid = {english}
}

@article{kessel1999Multiqubit,
  title = {Multiqubit Spin},
  author = {Kessel', Alexander R. and Ermakov, Vladimir L.},
  year = 1999,
  journal = {Journal of Experimental and Theoretical Physics Letters},
  volume = {70},
  pages = {61--65},
  publisher = {Springer},
  urldate = {2024-07-27}
}

@article{kessel2002Implementation,
  title = {Implementation Schemes in {{NMR}} of Quantum Processors and the {{Deutsch-Jozsa}} Algorithm by Using Virtual Spin Representation},
  author = {Kessel, Alexander R. and Yakovleva, Natalia M.},
  year = 2002,
  month = dec,
  journal = {Physical Review A},
  volume = {66},
  number = {6},
  pages = {062322},
  issn = {1050-2947, 1094-1622},
  doi = {10.1103/PhysRevA.66.062322},
  urldate = {2024-07-27},
  copyright = {http://link.aps.org/licenses/aps-default-license},
  langid = {english}
}

@article{kielpinski2002Architecture,
  title = {Architecture for a Large-Scale Ion-Trap Quantum Computer},
  author = {Kielpinski, David and Monroe, Chris and Wineland, David J.},
  year = 2002,
  journal = {Nature},
  volume = {417},
  number = {6890},
  pages = {709--711},
  publisher = {Nature Publishing Group UK London},
  urldate = {2025-05-20}
}

@article{kiktenko2025Colloquium,
  title = {{\emph{Colloquium}} : {{Qudits}} for Decomposing Multiqubit Gates and Realizing Quantum Algorithms},
  shorttitle = {{\emph{Colloquium}}},
  author = {Kiktenko, Evgeniy O. and Nikolaeva, Anastasiia S. and Fedorov, Aleksey K.},
  year = 2025,
  month = jun,
  journal = {Reviews of Modern Physics},
  volume = {97},
  number = {2},
  pages = {021003},
  issn = {0034-6861, 1539-0756},
  doi = {10.1103/RevModPhys.97.021003},
  urldate = {2025-08-29},
  langid = {english}
}

@article{leibfried2003Experimental,
  title = {Experimental Demonstration of a Robust, High-Fidelity Geometric Two Ion-Qubit Phase Gate},
  author = {Leibfried, Dietrich and DeMarco, Brian and Meyer, Volker and Lucas, David and Barrett, Murray and Britton, Joe and Itano, Wayne M. and Jelenkovi{\'c}, B. and Langer, Chris and Rosenband, Till},
  year = 2003,
  journal = {Nature},
  volume = {422},
  number = {6930},
  pages = {412--415},
  publisher = {Nature Publishing Group UK London},
  urldate = {2024-07-23}
}

@article{leibfried2003Quantum,
  title = {Quantum Dynamics of Single Trapped Ions},
  author = {Leibfried, D. and Blatt, R. and Monroe, C. and Wineland, D.},
  year = 2003,
  month = mar,
  journal = {Reviews of Modern Physics},
  volume = {75},
  number = {1},
  pages = {281--324},
  issn = {0034-6861, 1539-0756},
  doi = {10.1103/RevModPhys.75.281},
  urldate = {2024-07-23},
  copyright = {http://link.aps.org/licenses/aps-default-license},
  langid = {english}
}

@misc{loschnauer2024Scalable,
  title = {Scalable, High-Fidelity All-Electronic Control of Trapped-Ion Qubits},
  author = {L{\"o}schnauer, C. M. and Toba, J. Mosca and Hughes, A. C. and King, S. A. and Weber, M. A. and Srinivas, R. and Matt, R. and Nourshargh, R. and Allcock, D. T. C. and Ballance, C. J. and Matthiesen, C. and Malinowski, M. and Harty, T. P.},
  year = 2024,
  month = jul,
  number = {arXiv:2407.07694},
  eprint = {2407.07694},
  primaryclass = {physics, physics:quant-ph},
  publisher = {arXiv},
  urldate = {2024-07-27},
  archiveprefix = {arXiv}
}

@article{low2025Control,
  title = {Control and Readout of a 13-Level Trapped Ion Qudit},
  author = {Low, Pei Jiang and White, Brendan and Senko, Crystal},
  year = 2025,
  journal = {npj Quantum Information},
  volume = {11},
  number = {1},
  pages = {85},
  publisher = {Nature Publishing Group UK London},
  urldate = {2025-09-25}
}

@article{madsen2022Quantum,
  title = {Quantum Computational Advantage with a Programmable Photonic Processor},
  author = {Madsen, Lars S. and Laudenbach, Fabian and Askarani, Mohsen Falamarzi and Rortais, Fabien and Vincent, Trevor and Bulmer, Jacob FF and Miatto, Filippo M. and Neuhaus, Leonhard and Helt, Lukas G. and Collins, Matthew J.},
  year = 2022,
  journal = {Nature},
  volume = {606},
  number = {7912},
  pages = {75--81},
  publisher = {Nature Publishing Group UK London},
  urldate = {2024-07-27}
}

@article{magnus1954exponential,
  title = {On the Exponential Solution of Differential Equations for a Linear Operator},
  author = {Magnus, Wilhelm},
  year = 1954,
  month = nov,
  journal = {Communications on Pure and Applied Mathematics},
  volume = {7},
  number = {4},
  pages = {649--673},
  issn = {0010-3640, 1097-0312},
  doi = {10.1002/cpa.3160070404},
  urldate = {2024-07-27},
  langid = {english}
}

@article{main2025Distributed,
  title = {Distributed Quantum Computing across an Optical Network Link},
  author = {Main, D. and Drmota, P. and Nadlinger, D. P. and Ainley, E. M. and Agrawal, A. and Nichol, B. C. and Srinivas, R. and Araneda, G. and Lucas, D. M.},
  year = 2025,
  journal = {Nature},
  pages = {1--6},
  publisher = {Nature Publishing Group UK London},
  urldate = {2025-10-08}
}

@article{molmer1999Multiparticle,
  title = {Multiparticle {{Entanglement}} of {{Hot Trapped Ions}}},
  author = {M{\o}lmer, Klaus and S{\o}rensen, Anders},
  year = 1999,
  month = mar,
  journal = {Physical Review Letters},
  volume = {82},
  number = {9},
  pages = {1835--1838},
  issn = {0031-9007, 1079-7114},
  doi = {10.1103/PhysRevLett.82.1835},
  urldate = {2024-07-23},
  copyright = {http://link.aps.org/licenses/aps-default-license},
  langid = {english}
}

@article{monroe1995Demonstration,
  title = {Demonstration of a {{Fundamental Quantum Logic Gate}}},
  author = {Monroe, C. and Meekhof, D. M. and King, B. E. and Itano, W. M. and Wineland, D. J.},
  year = 1995,
  month = dec,
  journal = {Physical Review Letters},
  volume = {75},
  number = {25},
  pages = {4714--4717},
  issn = {0031-9007, 1079-7114},
  doi = {10.1103/PhysRevLett.75.4714},
  urldate = {2024-07-23},
  copyright = {http://link.aps.org/licenses/aps-default-license},
  langid = {english}
}

@article{monroe2013Scaling,
  title = {Scaling the {{Ion Trap Quantum Processor}}},
  author = {Monroe, C. and Kim, J.},
  year = 2013,
  month = mar,
  journal = {Science},
  volume = {339},
  number = {6124},
  pages = {1164--1169},
  issn = {0036-8075, 1095-9203},
  doi = {10.1126/science.1231298},
  urldate = {2024-07-23},
  langid = {english}
}

@article{moses2023RaceTrack,
  title = {A {{Race-Track Trapped-Ion Quantum Processor}}},
  author = {Moses, S. A. and Baldwin, C. H. and Allman, M. S. and Ancona, R. and Ascarrunz, L. and Barnes, C. and Bartolotta, J. and Bjork, B. and Blanchard, P. and Bohn, M. and Bohnet, J. G. and Brown, N. C. and Burdick, N. Q. and Burton, W. C. and Campbell, S. L. and Campora, J. P. and Carron, C. and Chambers, J. and Chan, J. W. and Chen, Y. H. and Chernoguzov, A. and Chertkov, E. and Colina, J. and Curtis, J. P. and Daniel, R. and DeCross, M. and Deen, D. and Delaney, C. and Dreiling, J. M. and Ertsgaard, C. T. and Esposito, J. and Estey, B. and Fabrikant, M. and Figgatt, C. and Foltz, C. and {Foss-Feig}, M. and Francois, D. and Gaebler, J. P. and Gatterman, T. M. and Gilbreth, C. N. and Giles, J. and Glynn, E. and Hall, A. and Hankin, A. M. and Hansen, A. and Hayes, D. and Higashi, B. and Hoffman, I. M. and Horning, B. and Hout, J. J. and Jacobs, R. and Johansen, J. and Jones, L. and Karcz, J. and Klein, T. and Lauria, P. and Lee, P. and Liefer, D. and Lu, S. T. and Lucchetti, D. and Lytle, C. and Malm, A. and Matheny, M. and Mathewson, B. and Mayer, K. and Miller, D. B. and Mills, M. and Neyenhuis, B. and Nugent, L. and Olson, S. and Parks, J. and Price, G. N. and Price, Z. and Pugh, M. and Ransford, A. and Reed, A. P. and Roman, C. and Rowe, M. and {Ryan-Anderson}, C. and Sanders, S. and Sedlacek, J. and Shevchuk, P. and Siegfried, P. and Skripka, T. and Spaun, B. and Sprenkle, R. T. and Stutz, R. P. and Swallows, M. and Tobey, R. I. and Tran, A. and Tran, T. and Vogt, E. and Volin, C. and Walker, J. and Zolot, A. M. and Pino, J. M.},
  year = 2023,
  month = dec,
  journal = {Physical Review X},
  volume = {13},
  number = {4},
  pages = {041052},
  issn = {2160-3308},
  doi = {10.1103/PhysRevX.13.041052},
  urldate = {2024-07-23},
  langid = {english}
}

@article{negnevitsky2018Repeated,
  title = {Repeated Multi-Qubit Readout and Feedback with a Mixed-Species Trapped-Ion Register},
  author = {Negnevitsky, Vlad and Marinelli, Matteo and Mehta, Karan K. and Lo, H.-Y. and Fl{\"u}hmann, Christa and Home, Jonathan P.},
  year = 2018,
  journal = {Nature},
  volume = {563},
  number = {7732},
  pages = {527--531},
  publisher = {Nature Publishing Group UK London},
  urldate = {2025-10-08}
}

@article{nikolaeva2022Decomposing,
  title = {Decomposing the Generalized {{Toffoli}} Gate with Qutrits},
  author = {Nikolaeva, A. S. and Kiktenko, E. O. and Fedorov, A. K.},
  year = 2022,
  month = mar,
  journal = {Physical Review A},
  volume = {105},
  number = {3},
  pages = {032621},
  issn = {2469-9926, 2469-9934},
  doi = {10.1103/PhysRevA.105.032621},
  urldate = {2024-07-27},
  langid = {english}
}

@article{nikolaeva2024Efficient,
  title = {Efficient Realization of Quantum Algorithms with Qudits},
  author = {Nikolaeva, Anastasiia S. and Kiktenko, Evgeniy O. and Fedorov, Aleksey K.},
  year = 2024,
  month = dec,
  journal = {EPJ Quantum Technology},
  volume = {11},
  number = {1},
  pages = {43},
  issn = {2662-4400, 2196-0763},
  doi = {10.1140/epjqt/s40507-024-00250-0},
  urldate = {2024-07-27},
  langid = {english}
}

@article{nikolaeva2024Universal,
  title = {Universal Quantum Computing with Qubits Embedded in Trapped-Ion Qudits},
  author = {Nikolaeva, Anastasiia S. and Kiktenko, Evgeniy O. and Fedorov, Aleksey K.},
  year = 2024,
  month = feb,
  journal = {Physical Review A},
  volume = {109},
  number = {2},
  pages = {022615},
  issn = {2469-9926, 2469-9934},
  doi = {10.1103/PhysRevA.109.022615},
  urldate = {2024-07-23},
  langid = {english}
}

@article{nikolaeva2025Scalable,
  title = {Scalable {{Improvement}} of the {{Generalized Toffoli Gate Realization Using Trapped-Ion-Based Qutrits}}},
  author = {Nikolaeva, Anastasiia S. and Zalivako, Ilia V. and Borisenko, Alexander S. and Semenin, Nikita V. and Galstyan, Kristina P. and Korolkov, Andrey E. and Kiktenko, Evgeniy O. and Khabarova, Ksenia Yu. and Semerikov, Ilya A. and Fedorov, Aleksey K. and Kolachevsky, Nikolay N.},
  year = 2025,
  month = aug,
  journal = {Physical Review Letters},
  volume = {135},
  number = {6},
  pages = {060601},
  issn = {0031-9007, 1079-7114},
  doi = {10.1103/p1z9-6w93},
  urldate = {2025-08-13},
  langid = {english}
}

@article{noiri2022Fast,
  title = {Fast Universal Quantum Gate above the Fault-Tolerance Threshold in Silicon},
  author = {Noiri, Akito and Takeda, Kenta and Nakajima, Takashi and Kobayashi, Takashi and Sammak, Amir and Scappucci, Giordano and Tarucha, Seigo},
  year = 2022,
  journal = {Nature},
  volume = {601},
  number = {7893},
  pages = {338--342},
  publisher = {Nature Publishing Group UK London},
  urldate = {2024-07-27}
}

@article{pino2021Demonstration,
  title = {Demonstration of the Trapped-Ion Quantum {{CCD}} Computer Architecture},
  author = {Pino, Juan M. and Dreiling, Jennifer M. and Figgatt, Caroline and Gaebler, John P. and Moses, Steven A. and Allman, M. S. and Baldwin, C. H. and {Foss-Feig}, Michael and Hayes, David and Mayer, Karl},
  year = 2021,
  journal = {Nature},
  volume = {592},
  number = {7853},
  pages = {209--213},
  publisher = {Nature Publishing Group UK London},
  urldate = {2024-07-23}
}

@article{pogorelov2021Compact,
  title = {Compact {{Ion-Trap Quantum Computing Demonstrator}}},
  author = {Pogorelov, I. and Feldker, T. and Marciniak, {\relax Ch}. D. and Postler, L. and Jacob, G. and Krieglsteiner, O. and Podlesnic, V. and Meth, M. and Negnevitsky, V. and Stadler, M. and H{\"o}fer, B. and W{\"a}chter, C. and Lakhmanskiy, K. and Blatt, R. and Schindler, P. and Monz, T.},
  year = 2021,
  month = jun,
  journal = {PRX Quantum},
  volume = {2},
  number = {2},
  pages = {020343},
  issn = {2691-3399},
  doi = {10.1103/PRXQuantum.2.020343},
  urldate = {2025-10-08},
  langid = {english}
}

@article{ralph2007Efficient,
  title = {Efficient {{Toffoli}} Gates Using Qudits},
  author = {Ralph, T. C. and Resch, K. J. and Gilchrist, A.},
  year = 2007,
  month = feb,
  journal = {Physical Review A},
  volume = {75},
  number = {2},
  pages = {022313},
  issn = {1050-2947, 1094-1622},
  doi = {10.1103/PhysRevA.75.022313},
  urldate = {2024-07-24},
  copyright = {http://link.aps.org/licenses/aps-default-license},
  langid = {english}
}

@article{ringbauer2022universal,
  title = {A Universal Qudit Quantum Processor with Trapped Ions},
  author = {Ringbauer, Martin and Meth, Michael and Postler, Lukas and Stricker, Roman and Blatt, Rainer and Schindler, Philipp and Monz, Thomas},
  year = 2022,
  journal = {Nature Physics},
  volume = {18},
  number = {9},
  pages = {1053--1057},
  publisher = {Nature Publishing Group UK London},
  urldate = {2024-07-23}
}

@article{roos2008Ion,
  title = {Ion Trap Quantum Gates with Amplitude-Modulated Laser Beams},
  author = {Roos, Christian F.},
  year = 2008,
  journal = {New Journal of Physics},
  volume = {10},
  number = {1},
  pages = {013002},
  publisher = {IOP Publishing},
  urldate = {2024-07-23}
}

@article{sawyer2021Wavelengthinsensitive,
  title = {Wavelength-Insensitive, Multispecies Entangling Gate for Group-2 Atomic Ions},
  author = {Sawyer, Brian C. and Brown, Kenton R.},
  year = 2021,
  journal = {Physical Review A},
  volume = {103},
  number = {2},
  pages = {022427},
  publisher = {APS},
  urldate = {2024-09-17}
}

@article{schafer2018Fast,
  title = {Fast Quantum Logic Gates with Trapped-Ion Qubits},
  author = {Sch{\"a}fer, V. M. and Ballance, C. J. and Thirumalai, K. and Stephenson, L. J. and Ballance, T. G. and Steane, A. M. and Lucas, D. M.},
  year = 2018,
  journal = {Nature},
  volume = {555},
  number = {7694},
  pages = {75--78},
  publisher = {Nature Publishing Group UK London},
  urldate = {2024-07-23}
}

@article{shapira2020Theory,
  title = {Theory of Robust Multiqubit Nonadiabatic Gates for Trapped Ions},
  author = {Shapira, Yotam and Shaniv, Ravid and Manovitz, Tom and Akerman, Nitzan and Peleg, Lee and Gazit, Lior and Ozeri, Roee and Stern, Ady},
  year = 2020,
  month = mar,
  journal = {Physical Review A},
  volume = {101},
  number = {3},
  pages = {032330},
  issn = {2469-9926, 2469-9934},
  doi = {10.1103/PhysRevA.101.032330},
  urldate = {2024-07-24},
  langid = {english}
}

@inproceedings{shor1994Algorithms,
  title = {Algorithms for Quantum Computation: Discrete Logarithms and Factoring},
  shorttitle = {Algorithms for Quantum Computation},
  booktitle = {Proceedings 35th Annual Symposium on Foundations of Computer Science},
  author = {Shor, Peter W.},
  year = 1994,
  pages = {124--134},
  publisher = {Ieee},
  urldate = {2024-07-27}
}

@article{smith2025SingleQubit,
  title = {Single-{{Qubit Gates}} with {{Errors}} at the 10 - 7 {{Level}}},
  author = {Smith, M. C. and Leu, A. D. and Miyanishi, K. and Gely, M. F. and Lucas, D. M.},
  year = 2025,
  month = jun,
  journal = {Physical Review Letters},
  volume = {134},
  number = {23},
  pages = {230601},
  issn = {0031-9007, 1079-7114},
  doi = {10.1103/42w2-6ccy},
  urldate = {2025-10-08},
  langid = {english}
}

@article{tan2015Multielement,
  title = {Multi-Element Logic Gates for Trapped-Ion Qubits},
  author = {Tan, Ting Rei and Gaebler, John P. and Lin, Yiheng and Wan, Yong and Bowler, R. and Leibfried, D. and Wineland, David J.},
  year = 2015,
  journal = {Nature},
  volume = {528},
  number = {7582},
  pages = {380--383},
  publisher = {Nature Publishing Group UK London},
  urldate = {2025-10-08}
}

@article{virtanen2020SciPy,
  title = {{{SciPy}} 1.0: Fundamental Algorithms for Scientific Computing in {{Python}}},
  shorttitle = {{{SciPy}} 1.0},
  author = {Virtanen, Pauli and Gommers, Ralf and Oliphant, Travis E. and Haberland, Matt and Reddy, Tyler and Cournapeau, David and Burovski, Evgeni and Peterson, Pearu and Weckesser, Warren and Bright, Jonathan},
  year = 2020,
  journal = {Nature methods},
  volume = {17},
  number = {3},
  pages = {261--272},
  publisher = {Nature Publishing Group},
  urldate = {2024-07-27}
}

@article{wan2019Quantum,
  title = {Quantum Gate Teleportation between Separated Qubits in a Trapped-Ion Processor},
  author = {Wan, Yong and Kienzler, Daniel and Erickson, Stephen D. and Mayer, Karl H. and Tan, Ting Rei and Wu, Jenny J. and Vasconcelos, Hilma M. and Glancy, Scott and Knill, Emanuel and Wineland, David J. and Wilson, Andrew C. and Leibfried, Dietrich},
  year = 2019,
  month = may,
  journal = {Science},
  volume = {364},
  number = {6443},
  pages = {875--878},
  issn = {0036-8075, 1095-9203},
  doi = {10.1126/science.aaw9415},
  urldate = {2025-10-08},
  langid = {english}
}

@article{wang2020Qudits,
  title = {Qudits and High-Dimensional Quantum Computing},
  author = {Wang, Yuchen and Hu, Zixuan and Sanders, Barry C. and Kais, Sabre},
  year = 2020,
  journal = {Frontiers in Physics},
  volume = {8},
  pages = {589504},
  publisher = {Frontiers Media SA},
  urldate = {2024-07-23}
}

@article{wang2022Fast,
  title = {Fast Multi-Qubit Global-Entangling Gates without Individual Addressing of Trapped Ions},
  author = {Wang, Kaizhao and Yu, Jing-Fan and Wang, Pengfei and Luan, Chunyang and Zhang, Jing-Ning and Kim, Kihwan},
  year = 2022,
  journal = {Quantum Science and Technology},
  volume = {7},
  number = {4},
  pages = {044005},
  publisher = {IOP Publishing},
  urldate = {2024-07-23}
}

@article{webb2018Resilient,
  title = {Resilient {{Entangling Gates}} for {{Trapped Ions}}},
  author = {Webb, A. E. and Webster, S. C. and Collingbourne, S. and Bretaud, D. and Lawrence, A. M. and Weidt, S. and Mintert, F. and Hensinger, W. K.},
  year = 2018,
  month = nov,
  journal = {Physical Review Letters},
  volume = {121},
  number = {18},
  pages = {180501},
  issn = {0031-9007, 1079-7114},
  doi = {10.1103/PhysRevLett.121.180501},
  urldate = {2024-07-25},
  langid = {english}
}

@article{wu2018Noise,
  title = {Noise Analysis for High-Fidelity Quantum Entangling Gates in an Anharmonic Linear {{Paul}} Trap},
  author = {Wu, Yukai and Wang, Sheng-Tao and Duan, L.-M.},
  year = 2018,
  month = jun,
  journal = {Physical Review A},
  volume = {97},
  number = {6},
  pages = {062325},
  issn = {2469-9926, 2469-9934},
  doi = {10.1103/PhysRevA.97.062325},
  urldate = {2024-07-23},
  langid = {english}
}

@article{wu2021Strong,
  title = {Strong {{Quantum Computational Advantage Using}} a {{Superconducting Quantum Processor}}},
  author = {Wu, Yulin and Bao, Wan-Su and Cao, Sirui and Chen, Fusheng and Chen, Ming-Cheng and Chen, Xiawei and Chung, Tung-Hsun and Deng, Hui and Du, Yajie and Fan, Daojin and Gong, Ming and Guo, Cheng and Guo, Chu and Guo, Shaojun and Han, Lianchen and Hong, Linyin and Huang, He-Liang and Huo, Yong-Heng and Li, Liping and Li, Na and Li, Shaowei and Li, Yuan and Liang, Futian and Lin, Chun and Lin, Jin and Qian, Haoran and Qiao, Dan and Rong, Hao and Su, Hong and Sun, Lihua and Wang, Liangyuan and Wang, Shiyu and Wu, Dachao and Xu, Yu and Yan, Kai and Yang, Weifeng and Yang, Yang and Ye, Yangsen and Yin, Jianghan and Ying, Chong and Yu, Jiale and Zha, Chen and Zhang, Cha and Zhang, Haibin and Zhang, Kaili and Zhang, Yiming and Zhao, Han and Zhao, Youwei and Zhou, Liang and Zhu, Qingling and Lu, Chao-Yang and Peng, Cheng-Zhi and Zhu, Xiaobo and Pan, Jian-Wei},
  year = 2021,
  month = oct,
  journal = {Physical Review Letters},
  volume = {127},
  number = {18},
  pages = {180501},
  issn = {0031-9007, 1079-7114},
  doi = {10.1103/PhysRevLett.127.180501},
  urldate = {2024-07-27},
  langid = {english}
}

@article{xue2022Quantum,
  title = {Quantum Logic with Spin Qubits Crossing the Surface Code Threshold},
  author = {Xue, Xiao and Russ, Maximilian and Samkharadze, Nodar and Undseth, Brennan and Sammak, Amir and Scappucci, Giordano and Vandersypen, Lieven MK},
  year = 2022,
  journal = {Nature},
  volume = {601},
  number = {7893},
  pages = {343--347},
  publisher = {Nature Publishing Group UK London},
  urldate = {2024-07-27}
}

@article{zalivako2025multiqudit,
  title = {Towards Multiqudit Quantum Processor Based on a \$\textasciicircum\textbraceleft 171\textbraceright\${{Yb}}\$\textasciicircum\textbraceleft +\textbraceright\$ Ion String: {{Realizing}} Basic Quantum Algorithms},
  shorttitle = {Towards Multiqudit Quantum Processor Based on a \$\textasciicircum\textbraceleft 171\textbraceright\${{Yb}}\$\textasciicircum\textbraceleft +\textbraceright\$ Ion String},
  author = {Zalivako, Ilia V. and Nikolaeva, Anastasiia S. and Borisenko, Alexander S. and Korolkov, Andrei E. and Sidorov, Pavel L. and Galstyan, Kristina P. and Semenin, Nikita V. and Smirnov, Vasilii N. and Aksenov, Mikhail A. and Makushin, Konstantin M. and Kiktenko, Evgeniy O. and Fedorov, Aleksey K. and Semerikov, Ilya A. and Khabarova, Ksenia Yu and Kolachevsky, Nikolay N.},
  year = 2025,
  month = apr,
  journal = {Quantum Reports},
  volume = {7},
  number = {2},
  eprint = {2402.03121},
  primaryclass = {quant-ph},
  pages = {19},
  issn = {2624-960X},
  doi = {10.3390/quantum7020019},
  urldate = {2025-05-20},
  archiveprefix = {arXiv}
}

@article{zhang2017Observation,
  title = {Observation of a Many-Body Dynamical Phase Transition with a 53-Qubit Quantum Simulator},
  author = {Zhang, Jiehang and Pagano, Guido and Hess, Paul W. and Kyprianidis, Antonis and Becker, Patrick and Kaplan, Harvey and Gorshkov, Alexey V. and Gong, Z.-X. and Monroe, Christopher},
  year = 2017,
  journal = {Nature},
  volume = {551},
  number = {7682},
  pages = {601--604},
  publisher = {Nature Publishing Group UK London},
  urldate = {2024-07-27}
}

@article{zhong2020Quantum,
  title = {Quantum Computational Advantage Using Photons},
  author = {Zhong, Han-Sen and Wang, Hui and Deng, Yu-Hao and Chen, Ming-Cheng and Peng, Li-Chao and Luo, Yi-Han and Qin, Jian and Wu, Dian and Ding, Xing and Hu, Yi and Hu, Peng and Yang, Xiao-Yan and Zhang, Wei-Jun and Li, Hao and Li, Yuxuan and Jiang, Xiao and Gan, Lin and Yang, Guangwen and You, Lixing and Wang, Zhen and Li, Li and Liu, Nai-Le and Lu, Chao-Yang and Pan, Jian-Wei},
  year = 2020,
  month = dec,
  journal = {Science},
  volume = {370},
  number = {6523},
  pages = {1460--1463},
  issn = {0036-8075, 1095-9203},
  doi = {10.1126/science.abe8770},
  urldate = {2024-07-27},
  langid = {english}
}

\appendix
\begin{widetext}
\section{Proof of factorizability of the MS gate evolution operator}
\label{sec:appendix_Stark}
In this section, we derive the commutator of the~$\hat{H}_{\mathrm{AC}}$ and~$\hat{H}_{\mathrm{MS}}$ Hamiltonians and prove that they commute if the Stark shifts of states~$\ket{0}$ and~$\ket{1}$ are equal. Using Eqs.~\eqref{eq:MSgate} and~\eqref{eq:ham_MS_AC}, we get
\begin{equation}
    \left[ \hat{H}_{\mathrm{AC}}, \hat{H}_{\mathrm{MS}}  \right]  = \sum_{l, j} \sum_k \sum_{s = 0}^{d-1} \hbar^2 \eta_{l, j} g_j(t)  \left( \hat{a}_{l} e^{- i \omega_{l } t} + \hat{a}_{l}^{\dagger} e^{ i \omega_{l} t} \right) \delta_{k,s}^{\mathrm{AC}}(t) \left[ \hat{\sigma}^{(j)}_{\phi_{S}}, \ketbra{s_k}{s_k} \right].
\end{equation}
Using the definition of~$\hat{\sigma}^{(j)}_{\phi_{S}}$ in \eqref{eq:sigmaphi}, we obtain
\begin{equation}
    \left[ \hat{\sigma}^{(j)}_{\phi_{S}},  \ketbra{s_k}{s_k} \right] = i \hat{\delta}_{jk} \hat{\sigma}^{(j)}_{\phi_{S} + \pi/2} \left( \hat{\delta}_{s0}    - \hat{\delta}_{s1} \right),
\end{equation}
where~$\hat{\delta}_{jk}$ is the Kronecker delta symbol. Then, the formula for the commutator can be simplified as follows:
\begin{equation}
     \left[ \hat{H}_{\mathrm{AC}}, \hat{H}_{\mathrm{MS}}  \right]  = \sum_{l, j} i \hbar^2 \eta_{l, j} g_j(t)  \left( \hat{a}_{l} e^{- i \omega_{l } t} + \hat{a}_{l}^{\dagger} e^{ i \omega_{l} t} \right) \hat{\sigma}^{(j)}_{\phi_{S} + \pi/2} (\delta_{j,0}^{\mathrm{AC}}(t) - \delta_{j,1}^{\mathrm{AC}}(t) ).
\end{equation}
Therefore, if~$\delta_{j,0}^{\mathrm{AC}}(t) \equiv \delta_{j,1}^{\mathrm{AC}}(t)$, the commutator is equal to zero.

\section{Multi-tone pulse shaping of the MS gate}\label{sec:appendix_AWG_shaping} 

In this Appendix we obtain the system, that we define in the main text in Eq.~\eqref{eq:SLSQP_system} and also discuss the important features of our multi-tone pulse shapes. The pulse function~$g(t)$ defined in Eq.~\eqref{eq:gFourier} can be seen as the superposition of~$P$ tones, therefore defined as the multi-tone pulse shapes.  
 The number of harmonics~$P$ and parameter~$n_{\min}$ define the range of detunings:
\begin{equation}\label{eq:gFourier}
\mu_{\min} = 2 \pi n_{\mathrm{min}} / \tau \leq \mu_p 
\leq \mu_{\max} = 2 \pi \left( n_{\mathrm{min}} + P-1\right) / \tau
\end{equation}
These parameters are selected to position all motional modes within the range $(\mu_{\min}, \mu_{\max})$, ensuring their efficient excitation. 

Then, using the known technique from Ref.~\cite{blumel2021Poweroptimal} we rewrite the decoupling condition~\eqref{eq:mode_decoupling} in terms of the amplitude vector~$\vec{\Omega}$:
\begin{equation}
\label{app_eq:decoupling_condition_shaping}
\alpha_{l, j}(\tau)=0, \quad l=1, .., N, \ j = 1,2 \Leftrightarrow \mathbf{M} \vec{\Omega}=0,
\end{equation}
where the matrix~$\mathbf{M}$ has the form
\begin{equation}
\label{app_eq:matrix_M_AWG}
\mathbf{M}_{l, p}=\int_0^\tau \sin \left(\mu_p t\right) \sin \left[\omega_l\left(\frac{\tau}{2}-t\right)\right] d t.
\end{equation}
The matrix~$\mathbf{M}$ is constructed by exploiting the anti-symmetry of the pulse function~$g(t)$, which reduces the decoupling conditions to the simplified form given in Eq.~\eqref{app_eq:gsin}. 
The decoupling conditions can be expressed as a system of linear equations in terms of the amplitude vector~$\vec{\Omega}$, solvable using standard linear algebra techniques. Lengthy but straightforward, one can obtain that the MS phases can be obtained in the form:
\begin{equation}
    \chi_{jk}^\mathrm{MS} = \vec{\Omega}^T \mathbf{D}_{jk}\vec{\Omega},
\end{equation}
where the matrices~$\mathbf{D}_{jk}$ are defined as
\begin{equation}
    \label{eq:matrix_D_AWG}
(\mathbf{D}_{jk})_{nm}  =\sum_{l} \eta_{l,j} \eta_{l,k} \int_0^\tau d t_1 \int_0^{t_1} d t_2 \\
 \sin(\mu_n t_2) \sin(\mu_m t_1) \sin \left[\omega_l\left(t_1-t_2\right)\right].
\end{equation}
 Since the~$\chi_{jk}^{\mathrm{MS}}$ is a scalar, the equation can be symmetrized using the matrix~ $ \mathbf{S}_{jk} = (\mathbf{D}_{jk} + \mathbf{D}_{jk}^T)/2$, see Eq.~\eqref{eq:chi_AWG}.
To find solutions non-sensitive in the first order to the motional modes drifts, we would also require that the derivatives of the phases with respect to secular frequencies be zero:
\begin{equation}
\label{eq:chi_linear_stabilization}
     \frac{\partial \chi_{jk}^{\mathrm{MS}}}{\partial \omega_l} =     \Vec{\Omega}^T  \mathbf{Q}_{jk, l}  \Vec{\Omega} = 0,
\end{equation}
for ions~$j, k = 1, 2$ and motional modes~$l = 1,..,N$. The matrices~$\mathbf{Q}_{jk, l}$ have the form:
\begin{equation}
\label{eq:C_projection}
    \mathbf{Q}_{jk, l} = \frac{\partial \mathbf{S}_{jk}}{\partial \omega_l}
\end{equation}
The stabilization conditions in Eq.~\eqref{eq:chi_linear_stabilization} are~$3N$ quadratic equations with multiple variables, which makes it difficult to find the solution numerically. Furthermore, some of the matrices~$\mathbf{Q}_{jk, l}$ may be positive or negative definite, so that the stabilization condition cannot be satisfied exactly. Instead, we employ the projection method introduced in Ref.~\cite{blumel2021Poweroptimal}. The idea of this method is to reduce the derivative~$\frac{\partial \chi_{jk}^{\mathrm{MS}}}{\partial \omega_l}$ by projecting out~$M$ vectors that express the highest sensitivity to motional frequency drifts. From Eq.~\eqref{eq:chi_linear_stabilization} we conclude that the vectors to be projected out are the eigenvectors of matrices~$\mathbf{Q}_{jk, l}$ with the highest magnitudes of their eigenvalues:
\begin{equation}
    \mathbf{Q}_{jk, l} \vec{V}_{jk, l}^{(m)} = \lambda_{jk, l}^{(m)} \vec{V}_{jk, l}^{(m)},\quad 1\leq m\leq M,
\end{equation}
where the eigenvalues~$\lambda_{jk, l}^{(m)}$ are sorted in descending order of their magnitude:~$|\lambda_{jk, l}^{(1)}| \geq |\lambda_{jk, l}^{(2)}| \geq .. \geq |\lambda_{jk, l}^{(P)}|$. The higher the eigenvalue~$\lambda_{jk, l}^{(m)}$, the more contribution of the vector~$\vec{V}_{jk, l}^{(m)}$ to the derivative in~\eqref{eq:chi_linear_stabilization}. Therefore, we project out the first~$M$ eigenvectors~$\vec{V}_{jk, l}^{(m)}$ from each matrix~$\mathbf{Q}_{jk, l}$. The final stabilization condition for phases~$\chi_{jk}^{\mathrm{MS}}$ can be written as
\begin{equation}
\label{eq:stabilization_condition}
    \mathbf{R}_{\chi}^T\vec{\Omega} = 0,
\end{equation}
where matrix~$\mathbf{R}_{\chi}$ is composed of the eigenvectors mentioned above as columns:
\begin{equation}
\label{eq:matrix_R}
    \mathbf{R}_{\chi} = \left[ \vec{V}_{jk, l}^{(m)} \right]_{(jk, l, m) \in \mathcal{I}},
\end{equation}
where~$\mathcal{I} = \big \{ (jk, l, m) \big \vert jk = \{11, 12, 22 \}, 1 \leq l \leq N, 1 \leq m \leq M   \big \}.$
The size of the matrix~$\mathbf{R}_\chi$ is~$P\times MN$ for each pair~$jk$.

In addition to stabilization of MS phases, we also implement~$K$-moment stabilization. This technique, employed in Ref.~\cite{blumel2021Poweroptimal} enhances the gate robustness by suppressing the residual entanglement between electronic and motional states. Similarly to stabilization of~$\chi_{jk}^{\mathrm{MS}}$ phases the moment stabilization tries to set the first~$K$ derivatives of the~$\alpha_{l, j}$ to zero:
\begin{equation}
    \frac{\partial^k \alpha_{l, j}}{\partial \omega_l^k}=0 \Leftrightarrow \mathbf{M}^{(k)} \Omega=0,
\end{equation}
where~$\mathbf{M}^{(k)} = \partial^k \mathbf{M} / \partial \omega_l^k$. Using this relations, we construct the condition for residual entanglement stabilization in a form analogous to Eq.~\eqref{eq:matrix_R}:
\begin{equation}
     \mathbf{R}_{\alpha}^T\vec{\Omega} = 0, \quad \mathbf{R}_{\alpha} = \left \{ \mathbf{M}^{(k)}, \mid k = 1, .. K \right \}.
\end{equation}
The complete stabilization matrix used in the system of equations~\eqref{eq:SLSQP_system} is then given by the concatenation of the latter matrices~$\mathbf{R} = [\mathbf{R}_{\alpha}; \mathbf{R}_{\chi}]$.

\section{Magnus expansion for MS and LS gates. Decoupling condition.} \label{sec:appendix_Magnus}
In our work we often use the Magnus expansion \cite{magnus1954exponential} to obtain the evolution operator of time dependent hamiltonian~$\hat{H}(t)$:
\begin{equation}
\label{app_eq:Magnus}
\hat{U}=\exp \left(\sum_{k = 1}^{\infty} \hat{A}_k(t)\right),
\end{equation}
where the first two operators are 
\begin{align}
& \hat{A}_1=-\frac{i}{\hbar} \int_0^t d t_1 \hat{H}\left(t_1\right)  \label{app_eq:Magnus_A1}\\
& \hat{A}_2=\frac{1}{2}\left(\frac{i}{\hbar}\right)^2 \int_0^t d t_1 \int_0^{t_1} d t_2\left[\hat{H}\left(t_1\right), \hat{H}\left(t_2\right)\right]
\label{app_eq:Magnus_A2}
\end{align}
Note, that the Magnus expansion is used both for the~$\hat{\sigma}_X \otimes \hat{\sigma}_X$ and the~$\hat{\sigma}_Z \otimes \hat{\sigma}_Z$ gates, as their Hamiltonians have a similar structure (compare Eqs.~\eqref{eq:MSgate} and~\eqref{eq:ham_LS_full}). For such Hamiltonians the Magnus expansion~\eqref{app_eq:Magnus} terminates at the second order, since all higher-order terms vanish due to the commutation relations~$[\hat{a}_l, [\hat{a}_l, \hat{a}_l^{\dag}]] = [\hat{a}_l^{\dag}, [\hat{a}_l, \hat{a}_l^{\dag}]] = 0$. 

This yields explicit expressions for the Magnus operators~$\hat{A}_1$ and~$\hat{A}_2$ for the MS gate that are used in the main text. We first derive the~$\hat{A}_1$:
\begin{equation}
    \label{app_eq:A_1}
    \hat{A}_1=\sum\limits_{l,j} \qty(\alpha_{l,j}\hat{a}^{\dagger}_l-\alpha^*_{l,j}\hat{a}_l)\hat{\sigma}^{(j)}_{X},
\end{equation}
where the complex coefficients~$\alpha_{l, j}$ define the displacement of the~$l^\mathrm{th}$ motional mode and~$j^\mathrm{th}$ ion. This parameters depend on the pulse function~$g(t)$ as was shown in Eq.~\eqref{eq:alpha}. Using linear transformations we can rewrite the system~\eqref{eq:alpha}:
 \begin{align}
 \int_0^{\tau} g(t) \cos \left[\omega_l\left(\frac{\tau}{2}-t\right)\right] d t &=0, \label{app_eq:gcos} 
 \\
 \int_0^{\tau} g(t) \sin \left[\omega_l\left(\frac{\tau}{2}-t\right)\right] d t &=0, \quad  ~\forall \ l = 1, \ldots, N. \label{app_eq:gsin} 
 \end{align}  
 For an anti-symmetric pulse function~$g(t)$, satisfying~$g(\tau - t) = -g(t)$ about the gate midpoint, the equations in~\eqref{app_eq:gcos} vanish identically. So only remaining equations~\eqref{app_eq:gsin} need to be solved. The second operator of the MS gate~\eqref{eq:A2} given in the main text follows directly from~\eqref{app_eq:Magnus_A2} when evaluated for the Hamiltionan of the MS gate~\eqref{eq:MSgate}. 
 
Similarly, we use Magnus expansion for LS gate and obtain the first operator~$\hat{A}_1$:
\begin{equation}
    \label{app_eq:A_1}
    \hat{A}_1=
    \sum\limits_{s = 0}^{d-1}  \sum\limits_{l,j} \qty(\alpha_{l,j, s}\hat{a}^{\dagger}_l-\alpha^*_{l,j, s}\hat{a}_l) \ketbra{s_j}{s_j},
\end{equation}
where the~$\alpha_{l, j, s}$ defines the displacement of the~$l^\mathrm{th}$ motional mode and the~$\ket{s}$ state of the~$j^\mathrm{th}$ qudit. The~$\alpha_{l, j, s}$ has the form:
\begin{equation}
    \alpha_{l, j, s} = -i\eta_{l,j}\int\limits_0^{\tau} \Omega_s(t) \cos (\mu t) e^{i\omega_l t} \,dt.
\end{equation}
Expressed in terms of the relative amplitudes~$\theta_s = \Omega_s/\Omega_0$, these parameters take the form~$\alpha_{l, j, s} = \theta_s \alpha_{l, j}$, where~$\alpha_{l, j} \equiv \alpha_{l, j, 0}$. This mirrors the structure of the~$\hat{\sigma}_X \otimes \hat{\sigma}_X$ gate coefficients from Eq.~\eqref{eq:alpha}.  

The second operator~$\hat{A}_2$ for the LS gate has the form
\begin{equation}
\hat{A}_2 = 
    i\sum_{s,s' = 0}^{d-1} \sum_{j, k = 1}^2 \chi_{jk;ss'}^{\mathrm{LS}} \ketbra{s_{j}}{s_{j}}\times \ketbra{s'_{k}}{s'_{k}}
\end{equation}
and can be obtained from the Magnus expansion formula~\eqref{app_eq:Magnus_A2} for the Hamiltonian of the LS gate~\eqref{eq:ham_LS_full}. The LS phases~$\chi_{jk}^\mathrm{LS}$ are defined in  the main text in Eq.~\eqref{eq:chi_LS_ss}.

\section{The existence of the solution for compensation of the single-qudit phases}\label{sec:appendix_existence}
The equation~\eqref{eq:chi_jk_MS} for~$\chi_{jk}^{\mathrm{MS}}$ in the main text  can be rewritten:
\begin{align}
\label{app_eq:chi_jk_theta}
\chi_{jk}^{\mathrm{MS}} &= \sum\limits_l   \eta_{l, j}  \eta_{l, k}  \Theta_{l}, \\ \Theta_{l}  &= \int\limits_0^{\tau}dt_1\int\limits_0^{t_1}
g(t_1) g(t_2) \sin(\omega_l(t_1-t_2))\,dt_2.
\label{app_eq:theta_l}
\end{align}
Here~$\theta_l$ depends on the pulse function~$g(t)$, gate duration~$\tau$ and 
\begin{equation}
    \label{app_eq:compensation_system_theta}
    \begin{cases}
        \chi_{12}^{\mathrm{MS}} = \sum\limits_l   \eta_{l, 1} \eta_{l, 2} \Theta_{l} = \chi, \\
     \chi_{jj}^{\mathrm{MS}} = \sum\limits_l   \eta_{l, j}^{2} \Theta_{l} = 0, \quad j = 1, 2.
    \end{cases}
\end{equation}
The solution of these system depends on the dimensionality of the~$\vec{\Theta}$ and the number of the independent constraints. The number of parameters~$\Theta_l$ matches the number of modes~$N$, while the number of constraints is at most three. When the target ions are chosen symmetrically to the center of the ion chain, this leads to~$\chi_{11}^\mathrm{MS} = \chi_{22}^\mathrm{MS}$, so there are only two independent constraints. To guarantee the existence of the solution, we need~$N \geq 2$, which can be achieved for any number of ions.

The key difference between multi-tone and amplitude modulations manifests in their ability to satisfy these compensation conditions. Given the solution~$\vec{\Theta}_0$ to the compensation system~\eqref{app_eq:compensation_system_theta}, we need to find a pulse function~$g(t)$, that yields~$\vec{\Theta} = \vec{\Theta}_0$ for~$\vec{\Theta}$ given in Eq.~\eqref{app_eq:theta_l}. Multi-tone modulation succeeds by placing the harmonics~$\mu_p$ near all motional modes, enabling arbitrary~$\Theta_l$ values. In contrast, the constant detuning~$\mu$ in amplitude modulation does not allow to achieve arbitrary~$\Theta_0$ because the laser field interact mostly with the spectrally nearest modes. However, when attempting to stabilize against fluctuations in motional mode frequencies, the projection method constrains the subspace of viable solutions. Consequently, we cannot guarantee that a stabilized solution exists across all parameters. As demonstrated in the main text, we succeed in finding the shapes for parameter values typically used in experimental settings. While our algorithm fails to converge in broader parameter regions, alternative optimizer configurations may yield solutions for specific experimental requirements.

\section{Native qudit gate representation of the cyclic shift~$\hat{X}_m^d$}\label{sec:appendix_X_m}
Any permutation of the qudit basis states can be performed by a series of resonant~$\pi$-pulses between different level pairs, since this operation is equivalent to a swap. Therefore, to define a specific permutation, one needs to only specify the indices of levels being swapped in a certain sequence. In a real setup, though, it is only viable to use one of the qudit states for swapping, usually because it is spectrally separated from the rest. In the following, we call this state~$\ket{0}$ or just 0 for brevity. Thus, the problem of cyclically shifting the whole basis by~$m$, accounting for the restriction made above, can be reformulated as follows: provide a sequence of indices~$1,2,\dots,d-1$ in an array of~$d$ elements with indices~$0,1,\dots,d-1$ to be successively swapped with index 0, such that the array gets shifted by~$m$ units at the end.

To solve the problem, we first assume that~$-d/2 < m\leq d/2$. If not, then redefine~$m\rightarrow (m\ \mathrm{mod}\ d) - d\cdot\theta\left((m\ \mathrm{mod}\ d)-d/2\right)$. This is valid, because~$\hat{X}_m^d = \hat{X}_{m+d}^d=\hat{X}_{-(d-m)}^d$. Second, any shift $\hat{X}_m^d$ with~$m<0$ can be performed by simply reversing the sequence for~$\hat{X}_{|m|}^d$, since pairwise swaps are their own inverses. Note that the sequence~$[1,2,\dots,d-1]$ realizes the shift~$\hat{X}_1^d$ in~$d-1=O(d)$ swaps. Technically, we can already build a working solution by applying the same unit shift~$m$ times, because~$\hat{X}_m^d=\left(\hat{X}_1^d\right)^m$. However, this would require~$O(md)$ native gates, which is undesirable for large~$m\lesssim d/2$: the total cost scaling would be~$O(d^2)$. In the following, we propose an algorithm that finds a suitable sequence with at most~$3d/2-2=O(d)$ swaps and with scaling independent of~$m$.

Our proposition is a variation of the known "juggling" algorithm, which is based on \emph{cycles}. Cycles are defined as index sequences of the form
\begin{equation}
a_n=(a_0+mn)\ \mathrm{mod}\ d,
\end{equation}
with every element being unique. In other words, a cycle represents a closed loop of indices that are separated by~$m$ from each other. A cyclic shift of the whole array by~$m$ will shift each cycle by 1. Taking into account these observations, the construction of the juggling algorithm is straightforward: identify all cycles, then shift each cycle by 1 (see Fig.~\ref{fig:appendix_X_m_orig}). 
\begin{figure}[h]
\centering
\begin{subfigure}[][][c]{.45\linewidth}
\centering
\caption{\label{fig:appendix_X_m_orig}}
\includegraphics[scale=0.9]{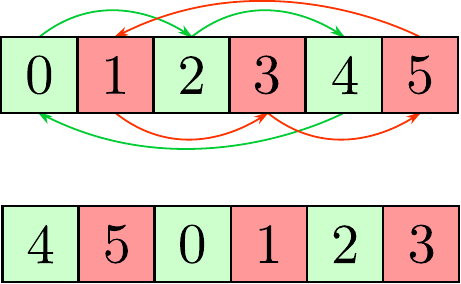}
\end{subfigure}
\begin{subfigure}[][][c]{.45\linewidth}
\centering
\caption{\label{fig:appendix_X_m_zeroswap}}
\includegraphics[scale=0.8]{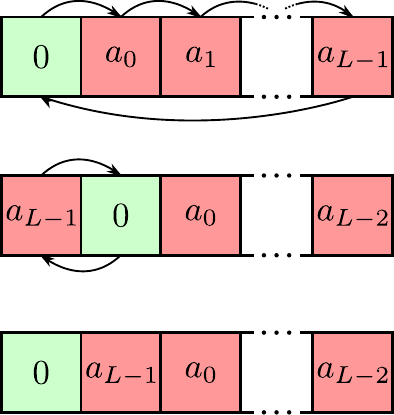}
\end{subfigure}
\caption{(a)~The original juggling algorithm for $d=6,m=2$. The arrows represent shifts within each cycle. (b)~Shifting an arbitrary cycle using element 0 as a buffer for swaps. Arrows represent shifts experienced by every element at the end of the respective cyclic rotation. The pairwise swaps still include 0.}
\end{figure}
In our case, we are restricted to only using the 0-th index for swapping. That means that only the cycle~${0,m,2m,\dots}$ can be shifted using the swap sequence~$[m,2m,\dots,d-m]$. For the rest of the cycles, we will use the 0-th element as a temporary buffer that will effectively prepend itself to the cycle. Then we will perform the same cyclic shift by 1, including the 0-th element. Afterwards, we swap the 0-th element with the first element of the cycle, achieving the desired shift (see Fig.~\ref{fig:appendix_X_m_zeroswap}). To identify all cycles, note that the cycle length (the number of distinct indices in a cycle) can be calculated as
\begin{equation}
L=\dfrac{\mathrm{lcm}(m,d)}{m}=\dfrac{d}{\mathrm{gcd}(m,d)},
\end{equation}
which is the same for all cycles. The number of cycles is therefore~$g=d/L=\mathrm{gcd}(m,d)$. Each index~$0,1,\dots,g-1$ belongs to a different cycle, because the separation between different indices in the same cycle has to be divisible by~$g$. Therefore, all cycles are generated by the first~$g$ indices. In conclusion, the swap sequence can be written as follows:
\begin{equation}
\begin{aligned}
[&m,2m,\dots,d-m,\\
&1,m+1,2m+1,\dots,d-m+1,1,\\
&2,m+2,2m+2,\dots,d-m+2,2,\\
&\dots,\\
&g-1,m+(g-1),2m+(g-1),\dots,d-m+(g-1),g-1].
\end{aligned}
\end{equation}
The number of swaps can be calculated exactly: the first cycle requires~$L-1$ swaps, the rest~$g-1$ cycles require~$L+1$ swaps because of the buffer index 0. Therefore, the total number of swaps is
\begin{equation}
N=(L-1) +(g-1)(L+1)=d+\mathrm{gcd}(m,d)-2.
\end{equation}
The highest value that~$\mathrm{gcd}(m,d)$ can take is~$d/2$, so the worst-case upper bound on the number of swaps is~$3d/2-2=O(d)$.

\section{Spin-echo sequence for embedded qubits}
\label{sec:spin-echo-embedded}
The spin-echo sequence of type~(c) comprises~$d^2/2$ two-particle gates, and thus its complexity increases rapidly with the qudit dimension~$d$. It is therefore desirable to identify scenarios in which this reduced gate configuration can be advantageously applied. As discussed in Sec.~\ref{sec:discussion}, such a gate enables the realization of two-qubit operations within embedded-qubit architectures. In the following, we provide a formal proof of this statement.

Consider implementing the light-shift (LS) gate on a pair of qudits labeled~$\mathsf{Q}_1$ and~$\mathsf{Q}_2$, each of dimension~$d$. We employ the mapping~$\ket{j}_{\mathsf{Q}} \leftrightarrow \ket{\mathrm{bin}_n(j)}_{\mathsf{q}_1 .. \mathsf{q}_n}$, which embed~$n$ qubits into a single qudit with~$d = 2^n$, and~$\mathrm{bin}_n(j)$ is the~$n$-binary representation of~$j$. 
% We would like to request that the notation "bin" be retained. We use "bin" to denote the n-bit binary representation of the integer, and this notation has also been adopted in previous APS publications, including Physical Review A 109, 022615 (2024) and Reviews of Modern Physics 97, 021003 (2025). Retaining this notation would ensure consistency with these references and with the existing literature.
For a system of two qudits, we denote the embedded qubits by~$\mathsf{q}_{ij}$, where the index~$i$ specifies the qudit~$\mathsf{Q}_i$ in which the qubit is encoded, and~$j$ labels the qubit itself. 
Our goal is to implement a two-qubit entangling operation of the form~$\hat{\sigma}_Z \otimes \hat{\sigma}_Z$: 
\begin{equation}
\label{app_eq:embedded_ZZ}
    \hat{\mathsf{ZZ}}_{\mathsf{q}_{1i}\mathsf{q}_{2j}} (\chi):= \exp(i \chi \hat{\sigma}_Z^{\mathsf{q}_i} \otimes \hat{\sigma}_Z^{\mathsf{q}_j}),
\end{equation}
Here,~$\hat{\sigma}_Z^{\mathsf{q}_i}$ denotes the Pauli-$Z$ operator acting on the qubit~$\mathsf{q}_i$. As a first step, we perform a transpilation of~$\hat{\sigma}_Z^{\mathsf{q}_i}$ into the corresponding qudit operators. Consider a~$2 \times 2$ unitary~$\hat{\mathcal{U}}$ acting within the subspace of the qubit~$\mathsf{q}_i$ with an identity operation acting in the remaining space of a qudit~\cite{nikolaeva2024Efficient}:
\begin{equation}
    \label{app_eq:U_qubit_embedded}
    \hat{\mathcal{U}}_{\mathsf{q}_i} = \sum_{(\alpha, \beta)} \hat{\mathcal{U}}^{\alpha, \beta},
\end{equation}
where~$\hat{\mathcal{U}}^{\alpha, \beta}$ means~$2\times 2 $ unitary that acts in the subspace of  the summation is made over all pairs~$(\alpha, \beta)$ that satisfy:
\begin{equation}
\label{app_eq:binary_alpha}
\begin{aligned}
    \mathrm{bin}_n(\alpha) &= x_1...x_{i-1}0x_{i+1}...x_{n}, \\
    \mathrm{bin}_n(\beta) &= x_1...x_{i-1}1x_{i+1}...x_{n},
\end{aligned}
\end{equation}
where~$x_1...x_{i-1}x_{i+1}...x_{n}$ are all possible bitstrings of length~$n-1$. These condition can be simplified
\begin{equation}
\label{app_eq:beta_alpha}
    \beta - \alpha = 2^{n-i}.
\end{equation}
The Eq.~\eqref{app_eq:U_qubit_embedded} can be used for~$\hat{\sigma}_Z$ gate. Consider the~$\hat{\sigma}_Z$ gate on the last qubit~$\mathsf{q}_n$, using the relation~\eqref{app_eq:beta_alpha}:
\begin{equation}
    \hat{\sigma}_Z^{\mathsf{q}_n} = \sum_{l = 0}^{d/2 - 1} \hat{\sigma}_Z^{2l, 2l + 1},
\end{equation}
where~$\hat{\sigma}_Z^{i, j} = \ketbra{i}{i} - \ketbra{j}{j}$. One can substitute these relation to Eq.~\eqref{app_eq:embedded_ZZ} for~$\hat{\mathsf{ZZ}}_{\mathsf{q}_{1n}\mathsf{q}_{2n}}$ and obtain:
\begin{equation}
    \hat{\mathsf{ZZ}}_{\mathsf{q}_{1n}\mathsf{q}_{2n}} (\chi) = \exp(i \chi \sum_{l, m = 0}^{d/2 - 1} \hat{\sigma}_Z^{2l, 2l + 1} \otimes \hat{\sigma}_Z^{2m, 2m + 1}).
\end{equation}
Since the operator inside the exponent is diagonal in the computational basis, we can obtain that states acquire the state-dependent phase:
\begin{equation}
     \hat{\mathsf{ZZ}}_{\mathsf{q}_{1n}\mathsf{q}_{2n}} (\chi) : \ket{ss'} \rightarrow e^{i \psi_{ss'}} \ket{ss'},
\end{equation}
where~$\psi_{ss'}$:
\begin{equation}
    \psi_{ss'} = \begin{cases}
        \chi,  & s - s' \mid 2,
        \\
        -\chi,  & s - s' \nmid 2.
    \end{cases}
\end{equation}
By comparison with the spin-echo sequence of type~(c), given in Eq.~\eqref{eq:phi_c}, it is evident that the two are equivalent up to a global phase. To realize the embedded~$\mathsf{ZZ}$-gate on other qubits, one can swap the qudits inside one qudit before and after the spin-echo sequence. Such swaps can be realized using only single-qudit gates. 

\end{widetext}
\end{document}